\documentclass[notitlepage,a4paper,aps,prd,onecolumn,superscriptaddress,nofootinbib,groupedaddress,longbibliography]{revtex4-2}
\usepackage{amsmath}
\usepackage{amsfonts}
\usepackage{amssymb}
\usepackage{accents}
\usepackage[utf8]{inputenc}
\usepackage[T1]{fontenc}
\usepackage{graphicx}
\usepackage[svgnames]{xcolor}
\usepackage[colorlinks=true]{hyperref}

\newcommand{\dd}{\mathrm{d}}

\newcommand{\vek}[1]{\underline{#1}}
\newcommand{\mat}[1]{\underline{\underline{#1}}}

\begin{document}

\title{Dynamical systems approach and cosmological attractors in newer general relativity}

\author{Manuel Hohmann}
\email{manuel.hohmann@ut.ee}
\affiliation{Laboratory of Theoretical Physics, Institute of Physics, University of Tartu, W. Ostwaldi 1, 50411 Tartu, Estonia}

\author{Ulbossyn Ualikhanova}
\email{ulbossyn.ualikhanova@gmail.com}
\affiliation{Department of General and Theoretical Physics, L.~N.~Gumilyov Eurasian National University, Satpayev Str. 2, 010008 Astana, Kazakhstan}

\begin{abstract}
We study the cosmological dynamics of a class of symmetric teleparallel gravity theories known as ``newer general relativity'' using the methods of dynamical systems, restricted to the case of vacuum solutions with a spatially flat Friedmann-Lemaître-Robertson-Walker metric. For the most general class of theories, we study generic properties of the solutions, in particular their fixed points, asymptotic behavior and effective dark energy. We then apply this approach to two phenomenologically motivated subclasses of theories, which we study in full detail. For these theories, we derive the complete space of solutions and cosmological attractors, which we display in a number of phase diagram. Depending on the particular theory at hand, we find different possible scenarios, including a turnaround followed by a big crunch, a big rip and an eternally expanding universe whose Hubble parameter asymptotically approaches zero. It turns that this different behavior can be explained by the effective dark energy barotropic index, which shows either phantom or non-phantom behavior, depending on the theory, but does not change dynamically between these two possibilities.
\end{abstract}

\maketitle


\section{Introduction}\label{sec:intro}
Despite its success across a wide range of scales~\cite{Baker:2014zba}, recent advances in observational cosmology have revealed significant problems with the standard cosmological model, known as the $\Lambda$CDM model, which combines general relativity (GR), a cosmological constant $\Lambda$ and cold dark matter (CDM). Inconsistencies such as the Hubble tension (the discrepancy between local and early measurements of the present-time Hubble parameter $H_0$) or the tensions in the measurements of the matter density parameter $\Omega_m$ and the structure growth rate $f\sigma_8$ and similar unresolved questions about the nature of dark energy and dark matter point to the need for theoretical frameworks that extend or modify GR~\cite{Planck:2018vyg,DiValentino:2021izs,DESI:2024mwx,DESI:2025zgx}. These observations have stipulated the development of numerous modified theories of gravity aimed at addressing these limitations~\cite{Nojiri:2006ri,Nojiri:2010wj,Faraoni:2010pgm,Clifton:2011jh,Bamba:2012cp,Nojiri:2017ncd,Bull:2015stt,Heisenberg:2018vsk,CANTATA:2021ktz,Odintsov:2023weg,Pfeifer:2023cgd}. Among these approaches, teleparallel gravity theories~\cite{BeltranJimenez:2019tjy,Hohmann:2022mlc} have received a high level of attention. These theories are built upon the assumption to describe gravity by a flat affine connection besides the metric, which may optionally be assumed to be either metric-compatible, leading to metric teleparallel gravity theories, or symmetric, leading to symmetric teleparallel gravity theories. In these theories, gravity is not attributed to the curvature of the Levi-Civita connection as in GR, but to the torsion or nonmetricity (or both) of the flat, teleparallel, affine connection.

In this article, we focus on the symmetric teleparallel class of theories~\cite{Nester:1998mp,Adak:2005cd,Mol:2014ooa,Adak:2018vzk}. The most prominent subclasses of symmetric teleparallel gravity include scalar-nonmetricity theories~\cite{Jarv:2018bgs,Runkla:2018xrv}, $f(Q)$ gravity~\cite{BeltranJimenez:2017tkd} and newer general relativity (Newer GR)~\cite{BeltranJimenez:2017tkd}. Here we focus on the latter. This class of theories is defined by the most general Lagrangian which is quadratic in the nonmetricity, and depends on five constant parameters, which govern the contribution of the five possible nonmetricity terms in the gravitational part of the action. This parameter space includes one point in which the field equations of the theory reduce to those of GR, whence this theory is known as the symmetric teleparallel equivalent of GR (STEGR). For the general class of Newer GR theories, numerous aspects have been studied, such as the propagation of gravitational waves~\cite{Hohmann:2018wxu}, their post-Newtonian limit~\cite{Flathmann:2020zyj}, a Hamiltonian formulation~\cite{DAmbrosio:2020nqu}, a pre-metric approach~\cite{Koivisto:2019ggr}, spherically symmetric solutions~\cite{Hohmann:2024phz} and the weak gravity limit~\cite{Golovnev:2024owe}, while only little attention has been paid to cosmology~\cite{Hohmann:2021ast}. These works have shown the existence of two one-parameter families of Newer GR theories, which are characterized by the consistency of their post-Newtonian limit with solar system observations and the absence of ghosts, and which are therefore of particular interest as viable gravitational theories.

An interesting property of the symmetric teleparallel, hence flat and symmetric, affine connection is the existence of a particular coordinate system, known as the coincident gauge, in which the connection coefficients vanish identically~\cite{BeltranJimenez:2022azb,Blixt:2023kyr}. In the early days of symmetric teleparallel cosmology, this coincident gauge has been assumed by default, while at the same time assuming a (usually flat) Friedmann-Lemaître-Robertson-Walker (FLRW) metric. However, this assumption neglects the fact that both the coincident gauge and the FLRW form of the metric require a particular choice of the coordinate system, and that there need not exist any coordinate system which simultaneously realizes both choices. In fact, it has been shown that there exist four branches of homogeneous and isotropic symmetric teleparallel geometries, three of which feature a flat FLRW metric, while the fourth branch describes the spatially curved case, and possesses the three spatially flat cases as different limits~\cite{Hohmann:2021ast}. For all three branches, the symmetric teleparallel cosmology is described by an additional function of time, in addition to the cosmological scale factor present in the FLRW metric (and the lapse function, which can be eliminated by making a fixed choice for the time coordinate).

A common framework to study the cosmological dynamics of any given model is the method of dynamical systems~\cite{Hohmann:2017jao}. It is based on expressing the cosmological field equations as a self-contained system of first-order differential equations, and allows to study different aspects of their dynamics. Among these, the existence and stability of fixed points is of particular interest, as they govern the possible asymptotic past and future behavior of the system. In the context of cosmology, these can be used to investigate the existence of a big bang and an accelerated late-time expansion of a given model, as well as possible future scenarios, including an eternal expansion (big freeze), turnaround followed by a re-collapse (big crunch) or another type of future finite time singularity. These different scenarios appear as attractive fixed points of the system, and can be found by charting its phase space. Further, studying the phase space also yields the possible trajectories connecting such fixed points, and thus possible cosmological histories.

The goal of this paper is to apply the method of dynamical systems to homogeneous and isotropic Newer GR cosmology, where we restrict ourselves to the three spatially flat branches, and consider only vacuum solutions. Both restrictions are motivated by simplicity, as they reduce the dimension of the resulting dynamical system. Further, they are also physically motivated. Observations show that the spatial curvature of the universe at the present time is very close to zero, which motivates considering a spatially flat FLRW geometry. Also the main aim of this article is to study attractor solutions for an expanding universe and the observed late-time accelerated of our universe, which shows that its dynamics is dominated by dark energy; this is in line with the fact that the contribution of matter in form of dust or radiation diminishes in an expanding universe due to their respective continuity equations, hence allowing us to neglect these matter contributions, in case there is any form of effective dark energy present, which takes the dominant role at late times.

The outline of this paper is as follows. In section~\ref{sec:nmcosmo}, we briefly review the symmetric teleparallel geometry, newer general relativity and its cosmological field equations. We use these to derive a formulation of the most general class of theories as a dynamical system in section~\ref{sec:general}, where we also study the generic properties of this system. We then turn our focus to two subclasses of theories, which are preferred by their observational properties and theoretical consistency, and analyze their cosmological dynamics in sections~\ref{sec:type1} and~\ref{sec:type2}, respectively. We summarize our results and provide an outlook in section~\ref{sec:conclusion}.

\section{Cosmology in newer general relativity}\label{sec:nmcosmo}
We begin with a brief review of symmetric teleparallel cosmology. The dynamical fields are a metric \(g_{\mu\nu}\) of Lorentzian signature and an independent affine connection with coefficients \(\Gamma^{\rho}{}_{\mu\nu}\), which is constrained by the conditions of vanishing curvature
\begin{equation}\label{eq:nocurv}
R^{\rho}{}_{\sigma\mu\nu} = \partial_{\mu}\Gamma^{\rho}{}_{\sigma\nu} - \partial_{\nu}\Gamma^{\rho}{}_{\sigma\mu} + \Gamma^{\rho}{}_{\lambda\mu}\Gamma^{\lambda}{}_{\sigma\nu} - \Gamma^{\rho}{}_{\lambda\nu}\Gamma^{\lambda}{}_{\sigma\mu} = 0\,,
\end{equation}
and vanishing torsion
\begin{equation}\label{eq:notors}
T^{\rho}{}_{\mu\nu} = \Gamma^{\rho}{}_{\nu\mu} - \Gamma^{\rho}{}_{\mu\nu} = 0\,,
\end{equation}
but in general possesses non-vanishing nonmetricity
\begin{equation}
Q_{\rho\mu\nu} = \nabla_{\rho}g_{\mu\nu}\,.
\end{equation}
The most general gravitational action functional which is quadratic in the nonmetricity is the so-called newer general relativity action~\cite{BeltranJimenez:2017tkd}
\begin{equation}\label{eq:action}
S_{\text{g}} = -\frac{1}{2\kappa^2}\int\dd^4x\sqrt{-g}(c_1Q^{\mu\nu\rho}Q_{\mu\nu\rho} + c_2Q^{\mu\nu\rho}Q_{\rho\mu\nu} + c_3Q^{\rho\mu}{}_{\mu}Q_{\rho\nu}{}^{\nu} + c_4Q^{\mu}{}_{\mu\rho}Q_{\nu}{}^{\nu\rho} + c_5Q^{\mu}{}_{\mu\rho}Q^{\rho\nu}{}_{\nu})\,,
\end{equation}
where \(\kappa^2\) denotes the gravitational constant and \(c_1, \ldots, c_5\) are constant parameters, whose values select a particular theory. Among their possible values, there are two one-parameter families of theories which are distinguished by the fact that their post-Newtonian limit agrees with that of general relativity~\cite{Flathmann:2020zyj} and there are no ghosts in a linear expansion around a flat Minkowski background~\cite{Bello-Morales:2024vqk}. These theories can be parametrized as~\cite{Hohmann:2024phz}
\begin{equation}\label{eq:ngrtype1}
c_1 = -\frac{1}{4}\,, \quad
c_2 = \frac{\epsilon}{4} + \frac{1}{2}\,, \quad
c_3 = \frac{1}{4}\,, \quad
c_4 = -\frac{\epsilon}{4}\,, \quad
c_5 = -\frac{1}{2}\,,
\end{equation}
denoted type 1, as well as
\begin{equation}\label{eq:ngrtype2}
c_1 = -\frac{1}{4}\,, \quad
c_2 = \frac{1}{2}\,, \quad
c_3 = \frac{\epsilon}{2} + \frac{1}{4}\,, \quad
c_4 = 0\,, \quad
c_5 = -\frac{1}{2}\,,
\end{equation}
denoted type 2, where \(\epsilon\) is a free parameter. Note that for \(\epsilon = 0\), both types of theories take the common limit
\begin{equation}\label{eq:ngrstegr}
c_1 = -\frac{1}{4}\,, \quad
c_2 = \frac{1}{2}\,, \quad
c_3 = \frac{1}{4}\,, \quad
c_4 = 0\,, \quad
c_5 = -\frac{1}{2}
\end{equation}
of the symmetric teleparallel equivalent of general relativity (STEGR).

In this article, we study the cosmological dynamics of newer general relativity. For this purpose, we consider a Friedmann-Lemaître-Robertson-Walker metric, which we decompose as
\begin{equation}
g_{\mu\nu} = -n_{\mu}n_{\nu} + h_{\mu\nu}
\end{equation}
into the hypersurface conormal \(n_{\mu}\) and spatial metric \(h_{\mu\nu}\), whose non-vanishing components are given by
\begin{equation}
n_t = -N\,, \quad
h_{rr} = \frac{A^2}{\chi^2}\,, \quad
h_{\vartheta\vartheta} = A^2r^2\,, \quad
h_{\varphi\varphi} = h_{\vartheta\vartheta}\sin^2\vartheta
\end{equation}
in spherical coordinate \((t, r, \vartheta, \varphi)\), where we used the abbreviation \(\chi = \sqrt{1 - kr^2}\), where \(k \in \mathbb{R}\) indicates the spatial curvature, and \(N(t)\) and \(A(t)\) denote the lapse function and scale factor, respectively. With the help of these quantities, the most general homogeneous and isotropic form of the nonmetricity, which in turn defines the teleparallel connection, is given by~\cite{Hohmann:2021ast}
\begin{equation}\label{eq:cosmononmet}
Q_{\rho\mu\nu} = 2Q_1n_{\rho}n_{\mu}n_{\nu} + 2Q_2n_{\rho}h_{\mu\nu} + 2Q_3h_{\rho(\mu}n_{\nu)}
\end{equation}
in terms of three functions \(Q_1(t), Q_2(t), Q_3(t)\). In order to maintain the flatness of the connection, these are not arbitrary, but must fall into one of four branches. Introducing the Hubble parameter
\begin{equation}
H = \frac{\partial_tA}{NA} = \frac{\mathcal{L}_{n}A}{A} = \frac{\dot{A}}{A}\,,
\end{equation}
we have for \(k = 0\) the first branch
\begin{equation}
Q_1 = K\,, \quad
Q_2 = -H\,, \quad
Q_3 = 0\,,
\end{equation}
the second branch
\begin{equation}
Q_1 = 2H + \frac{\dot{K}}{K}\,, \quad
Q_2 = -H\,, \quad
Q_3 = K
\end{equation}
and the third branch
\begin{equation}
Q_1 = -K - \frac{\dot{K}}{K}\,, \quad
Q_2 = K - H\,, \quad
Q_3 = K\,,
\end{equation}
while for \(k \neq 0\) there is a unique branch
\begin{equation}
Q_1 = -K - \frac{\dot{K}}{K}\,, \quad
Q_2 = K - H\,, \quad
Q_3 = K + \frac{k}{KA^2}\,.
\end{equation}
In all three cases, the cosmological dynamics of the teleparallel connection is described by a single function \(K(t)\). Inserting the cosmologically symmetric geometry into the gravitational field equations derived from the action~\eqref{eq:action}, one finds that the theory parameters enter into the cosmological field equations only in the four linear combinations
\begin{equation}\label{eq:cosmopar}
a_1 = 2(c_1 + 3c_3)\,, \quad
a_2 = 2(2c_3 + c_5)\,, \quad
a_3 = 2(c_1 + c_2 + c_3 + c_4 + c_5)\,, \quad
a_4 = 2(c_2 - c_4 + c_5)\,,
\end{equation}
and so it is convenient to use these instead of the original parametrization. With this definition, and complementing the gravitational part of the field equations with a homogeneous and isotropic energy-momentum tensor describing a perfect fluid with density \(\rho\) and pressure \(p\), we find that the dynamics of the first branch become
\begin{subequations}\label{eq:cosmo1}
\begin{align}
2a_3\dot{K} + 3a_2\dot{H} + a_3K^2 + 3(a_2 + 2a_3)HK + 3(a_1 + 3a_2)H^2 &= \kappa^2\rho\,,\\
-a_2\dot{K} - 2a_1\dot{H} + a_3K^2 - 3a_1H^2 &= \kappa^2p\,,
\end{align}
\end{subequations}
while for the second branch we have
\begin{subequations}\label{eq:cosmo2}
\begin{align}
2a_3\frac{\ddot{K}}{K} - a_3\frac{\dot{K}^2}{K^2} + (3a_2 + 10a_3)\frac{H\dot{K}}{K} + (3a_2 + 4a_3)\dot{H} + \frac{9}{4}(a_2 - 2a_3 + a_4)K^2 + 3a_4HK + (3a_1 + 15a_2 + 16a_3)H^2 &= \kappa^2\rho\,,\\
-a_2\frac{\ddot{K}}{K} + (a_2 + a_3)\frac{\dot{K}^2}{K^2} + 4a_3\frac{H\dot{K}}{K} + a_4\dot{K} - 2(a_1 + a_2)\dot{H} + \frac{a_2 - 2a_3 + a_4}{4}K^2 + 2a_4HK + (4a_3 - 3a_1)H^2 &= \kappa^2p\,,
\end{align}
\end{subequations}
the third branch yields
\begin{subequations}\label{eq:cosmo3}
\begin{align}
-2a_3\frac{\ddot{K}}{K} + 3a_3\frac{\dot{K}^2}{K^2} - 3(a_2 + 2a_3)\frac{H\dot{K}}{K} + 3a_2\dot{H} + \frac{21a_2 + 22a_3 - 3a_4}{4}K^2 - 3(6a_2 + 4a_3 - a_4)HK + 3(a_1 + 3a_2)H^2 &= \kappa^2\rho\,,\\
a_2\frac{\ddot{K}}{K} + (a_3 - a_2)\frac{\dot{K}^2}{K^2} + (6a_2 + 4a_3 - a_4)\dot{K} - 2a_1\dot{H} + \frac{21a_2 + 22a_3 - 3a_4}{4}K^2 - 3a_1H^2 &= \kappa^2p\,,
\end{align}
\end{subequations}
and finally the (spatially curved) fourth branch yields
\begin{subequations}\label{eq:cosmo4}
\begin{align}
-2a_3\frac{\ddot{K}}{K} + 3a_3\frac{\dot{K}^2}{K^2} - 3(a_2 + 2a_3)\frac{H\dot{K}}{K} + 3a_2\dot{H} + \frac{21a_2 + 22a_3 - 3a_4}{4}K^2 - 3(6a_2 + 4a_3 - a_4)HK &\nonumber\\
\phantom{0}+ 3(a_1 + 3a_2)H^2 + 3a_4k\frac{H}{A^2K} + \frac{3}{2}k\frac{2a_1 - a_2 - 4a_3 + a_4}{A^2} + \frac{9}{4}k^2\frac{a_2 - 2a_3 + a_4}{A^4K^2} &= \kappa^2\rho\,,\\
a_2\frac{\ddot{K}}{K} + (a_3 - a_2)\frac{\dot{K}^2}{K^2} + (6a_2 + 4a_3 - a_4)\dot{K} - 2a_1\dot{H} + \frac{21a_2 + 22a_3 - 3a_4}{4}K^2 - 3a_1H^2 &\nonumber\\
\phantom{0}- a_4k\frac{\dot{K}}{A^2K^2}- \frac{1}{2}k\frac{2a_1 - a_2 - 4a_3 + a_4}{A^2} + \frac{1}{4}k^2\frac{a_2 - 2a_3 + a_4}{A^4K^2} &= \kappa^2p\,.
\end{align}
\end{subequations}
In the remainder of this article, we will only consider the first three branches, which correspond to the spatially flat case \(k = 0\). Further, we will restrict ourselves to the vacuum dynamics of these theories, and thus set \(\rho = p = 0\).

\section{Dynamical systems approach and generic features}\label{sec:general}
We now derive the general cosmological dynamics of newer general relativity and derive a few generic properties, with the following restrictions. First, we will focus on the case of vacuum cosmology only, and therefore assume vanishing matter variables \(\rho = p = 0\). Further, we will not discuss the fourth branch~\eqref{eq:cosmo4} corresponding to a spatially curved Friedmann-Lemaître-Robertson-Walker universe, and restrict ourselves to the spatially flat branches. Both restrictions are physically motivated by our aim to study attractor solutions for an expanding universe, in which we may assume that both matter and spatial curvature can be neglected, and the expansion is dominated by the gravitational field variables. They also serve the purpose of reducing the number of variables in the resulting dynamical system, by eliminating the matter variables \(\rho, p\) and the scale factor \(A\) from the cosmological equations. We start by formulating the cosmological field equations for generic values of the theory parameters as a dynamical system in section~\ref{ssec:dynsysg}. It turns out that this approach excludes a particular subset of parameter values, in which the system becomes degenerate, and which we discuss in detail in section~\ref{ssec:degeng}. We then proceed with the non-degenerate system. We find that it can be decomposed into two parts, which we study in section~\ref{ssec:prodecg}. This decomposition will be useful for our discussion of fixed points in section~\ref{ssec:fixpg}, and essential for determining the asymptotic behavior of solutions in section~\ref{ssec:profixg}, which constitutes the main result of this article. The latter is further supported by the effective dark energy description in section~\ref{ssec:effdarkeng}.

\subsection{Dynamical system formulation}\label{ssec:dynsysg}
The starting point for our derivation are the cosmological field equations detailed in the previous section. In order to make use of the method of dynamical systems, we first need to formulate the cosmological dynamics as a system of first-order ordinary differential equations. While this is already the case for the first branch~\eqref{eq:cosmo1}, which becomes
\begin{subequations}\label{eq:dynsys1}
\begin{align}
2a_3\dot{K} + 3a_2\dot{H} + a_3K^2 + 3(a_2 + 2a_3)HK + 3(a_1 + 3a_2)H^2 &= 0\,,\label{eq:dynsys11}\\
-a_2\dot{K} - 2a_1\dot{H} + a_3K^2 - 3a_1H^2 &= 0\label{eq:dynsys12}
\end{align}
\end{subequations}
under the assumption of vacuum cosmology, we see that the other two branches~\eqref{eq:cosmo2} and~\eqref{eq:cosmo3} are of second order due to the presence of \(\ddot{K}\). In order to obtain a first-order system, we therefore need to replace \(\dot{K}\) by a new dynamical variable. For this purpose, note that \(\dot{K}\) always appears in the form
\begin{equation}
L = \frac{\dot{K}}{K} = \frac{\dd}{\dd t}\ln K\,,
\end{equation}
which must remain finite in order for the dynamics to be well-defined, i.e., the case \(K = 0\) with \(\dot{K} \neq 0\) must be excluded. This can be achieved by choosing \(L\) as above as new variable, so that we can substitute
\begin{subequations}
\begin{align}
\dot{K} &= KL\,,\\\label{eq:vardef}
\ddot{K} &= \dot{K}L + K\dot{L} = K(L^2 + \dot{L})\,.
\end{align}
\end{subequations}
The first of these equations is the desired first-order differential equation governing the dynamics of \(K\) as a function of \(K\) and \(L\), while the dynamics of \(H\) and \(L\) is governed by the equations
\begin{subequations}\label{eq:dynsys2}
\begin{align}
(3a_2 + 4a_3)\dot{H} + 2a_3\dot{L} + a_3L^2 + (3a_2 + 10a_3)HL + \frac{9}{4}(a_2 - 2a_3 + a_4)K^2 + 3a_4HK + (3a_1 + 15a_2 + 16a_3)H^2 &= 0\,,\label{eq:dynsys21}\\
-2(a_1 + a_2)\dot{H} - a_2\dot{L} + a_3L^2 + 4a_3HL + a_4KL + \frac{a_2 - 2a_3 + a_4}{4}K^2 + 2a_4HK + (4a_3 - 3a_1)H^2 &= 0\label{eq:dynsys22}
\end{align}
\end{subequations}
for the second branch and
\begin{subequations}\label{eq:dynsys3}
\begin{align}
3a_2\dot{H} - 2a_3\dot{L} + a_3L^2 - 3(a_2 + 2a_3)HL + \frac{21a_2 + 22a_3 - 3a_4}{4}K^2 - 3(6a_2 + 4a_3 - a_4)HK + 3(a_1 + 3a_2)H^2 &= 0\,,\label{eq:dynsys31}\\
-2a_1\dot{H} + a_2\dot{L} + a_3L^2 + (6a_2 + 4a_3 - a_4)KL + \frac{21a_2 + 22a_3 - 3a_4}{4}K^2 - 3a_1H^2 &= \label{eq:dynsys32}0
\end{align}
\end{subequations}
for the third branch.

The next step consists in solving the equations~\eqref{eq:dynsys1}, \eqref{eq:dynsys2} and~\eqref{eq:dynsys3}, along with~\eqref{eq:vardef}, for the first order derivatives of the dynamical variables. This step is greatly simplified that these equations turn out to form a linear system of equations in these quantities, which are \(\dot{H}\) and \(\dot{K}\) in the first branch, and in addition also \(\dot{L}\) in the other two branches. However, care must be taken in the case that this system becomes degenerate, which means that one of these derivatives can be eliminated from the dynamical equations, and a constraint among the variables \(H\) and \(K\) in the first branch, or in addition also \(L\) in the other two branches is obtained. It turns out that the system is non-degenerate if and only if
\begin{equation}\label{eq:nondegcond}
4a_1a_3 - 3a_2^2 \neq 0
\end{equation}
in any of the three branches. In this case, we can solve the system for the first order derivatives. For the first branch~\eqref{eq:dynsys1}, we find the system
\begin{subequations}\label{eq:nondeg1}
	\begin{align}
		\dot{H} &= \frac{3 \left(a_1a_2 + 3 a_2^2- 2 a_1 a_3\right) H^2 + 3 a_2 \left(a_2 + 2 a_3\right) H K + a_3 \left(a_2 + 2 a_3\right) K^2}{4 a_1 a_3-3 a_2^2 }\,,\\
		\dot{K} &= \frac{3 a_1 \left(2 a_1 + 3 a_2\right) H^2 + 6 a_1 \left(a_2 + 2 a_3\right) H K + \left(2 a_1 + 3 a_2\right) a_3 K^2}{3 a_2^2 - 4 a_1 a_3}\,,
	\end{align}
\end{subequations}
while for the second branch~\eqref{eq:dynsys2} we have
\begin{subequations}\label{eq:nondeg2}
	\begin{align}
		\dot{H} &= \frac{1}{4 \left( 4 a_1 a_3 - 3 a_2^2 \right)} \Bigg[
		12 a_1 \left(a_2 - 2 a_3\right) H^2
		+ a_2 \left(3 a_4 K \left(4 H + 3 K\right)
		+ 4 a_3 \left(\left(2 H + L\right) \left(8 H + L\right) - 4K^2\right)\right)
		\nonumber \\
		& \quad + 3 a_2^2 \left(3 K^2 + 4 H \left(5 H + L\right)\right)
		+ 2 a_3 \left(a_4 K \left(8 H + K + 4 L\right)
		+ 2 a_3 \left(2 \left(2H +L\right)^2 - K^2\right)\right)
		\Bigg], \\
		\dot{K} &= K L,  \\
		\dot{L} &= \frac{1}{4 \left(3 a_2^2 - 4 a_1 a_3\right)} \Bigg[
		24 a_1^2 H^2
		+ 4 a_3 \left(a_4 K \left(8 H + K + 4 L\right)
		+ 2 a_3 \left(2 \left(2H + L\right)^2 - K^2\right)\right)
		\nonumber \\
		& \quad + 3 a_2^2 \left(7 K^2 + 40 H^2 + 8HL\right)
		+ a_2 \left(3 a_4 K \left(16 H + 7 K + 4 L\right)
		- 38 a_3 K^2 + 4 a_3 \left(2 H + L\right) \left(22 H + 5L\right)\right)
		\nonumber \\
		& \quad + 2 a_1 \Big(3 a_4 K \left(4 H + 3 K\right)
		+ 3 a_2 \left(18 H^2 + 3 K^2 + 4 H L\right)
		+ 2 a_3 \left(20 H^2 - 9 K^2 + 20 H L + 2 L^2\right)\Big)
		\Bigg].
	\end{align}
\end{subequations}
and finally the third branch~\eqref{eq:dynsys3} takes the form
\begin{subequations}\label{eq:nondeg3}
	\begin{align}
		\dot{H} &= \frac{1}{4 \left(3 a_2^2 - 4 a_1 a_3\right)} \Bigg[
		-12 a_1 \left(a_2 - 2 a_3\right) H^2
		+ a_2 \left(3 a_4 K \left(-4 H + K\right)
		- 4 a_3 \left(16 K^2 + 12 K L + L^2 - 6 H \left(2 K + L\right)\right)\right)
		\nonumber \\
		& \quad - 3 a_2^2 \left(12 H^2 + 7 K^2 - 4 H \left(6 K + L\right)\right)
		+ 2 a_3 \left(a_4 K \left(3 K + 4 L\right)
		- 2 a_3 \left(11 K^2 + 8 K L + 2 L^2\right)\right)
		\Bigg], \\
		\dot{K} &= K L,  \\
		\dot{L} &= \frac{1}{4 \left(3 a_2^2 - 4 a_1 a_3\right)} \Bigg[
		-24 a_1^2 H^2
		+ 3 a_2 \left((3 a_4 - 21 a_2 - 22 a_3) K^2
		+ 4 (a_4 - 6 a_2 - 4 a_3) K L - 4 a_3 L^2\right)
		\nonumber \\
		& \quad + a_1 \Big(6 a_4 K \left(-4 H + K\right)
		+ a_3 \left(-44 K^2 - 8 L^2 + 48 H \left(2 K + L\right)\right)
		- 6 a_2 \left(6 H^2 + 7 K^2 - 4 H \left(6 K + L\right)\right)\Big)
		\Bigg].
	\end{align}
\end{subequations}
which shows the importance of the condition~\eqref{eq:nondegcond} due to its appearance in the denominator.

\subsection{Degenerate cases}\label{ssec:degeng}
Before studying the generic properties of the dynamical system derived in the previous section, we take a closer look at the degenerate case, which is obtained if the parameters satisfy the condition
\begin{equation}\label{eq:degencond}
4a_1a_3 - 3a_2^2 = 0\,.
\end{equation}
The space of theories satisfying this condition can conveniently be parametrized by introducing two parameters \(\alpha, C\), in terms of which the three parameters \(a_{1,2,3}\) are expressed as
\begin{equation}\label{eq:degenpar}
a_1 = C\cos^2\alpha\,, \quad
a_2 = \frac{C\sin 2\alpha}{\sqrt{3}}\,, \quad
a_3 = C\sin^2\alpha\,.
\end{equation}
The advantage of choosing this parametrization will become apparent below, when we discuss the dynamics of the degenerate systems.

\subsubsection{Branch 1}\label{sssec:degeng1}
We start with the first branch, where we will assume \(C \neq 0\), since otherwise the dynamical equations will trivially vanish. Inserting the parametrization~\eqref{eq:degenpar} and dividing by \(C\), the dynamical equations~\eqref{eq:dynsys1} become
\begin{subequations}\label{eq:degen10}
	\begin{align}
		3 H^2 \cos^2{\alpha} + 2 \dot{K} \sin^2{\alpha} + K (6H + K) \sin^2{\alpha} + \sqrt{3} \dot{H} \sin{2\alpha} + \sqrt{3} H (3H + K) \sin{2\alpha}&=0\,,\\
		- 2 \dot{H} \cos^2{\alpha} - 3 H^2 \cos^2{\alpha} + K^2 \sin^2{\alpha} - \frac{\dot{K} \sin{2\alpha}}{\sqrt{3}}&= 0\,.
	\end{align}
\end{subequations}
It follows that there exists a unique linear combination of these equations such that the derivative terms cancel, and one obtains the constraint
\begin{equation}\label{eq:degcons1}
\left(\cos\alpha + \sqrt{3}\sin\alpha\right)\left(\sqrt{3}H\cos\alpha + K\sin\alpha\right)^2 = 0\,.
\end{equation}
Two cases need to be considered:
\begin{enumerate}
\item
We first consider the case that the first factor is non-vanishing. In this case we find the constraint
\begin{equation}
\sqrt{3}H\cos\alpha + K\sin\alpha = 0\,,
\end{equation}
which possesses the general solution
\begin{equation}
H = X\sin\alpha\,, \quad
K = -\sqrt{3}X\cos\alpha\,,
\end{equation}
where \(X = X(t)\) is the only dynamical variable in this case. Inserting this solution, along with its time derivative, into the original equations~\eqref{eq:degen10}, we find that these are now satisfied identically. Hence, \(X\) remains undetermined.
\item
We then consider the case that the first factor in the constraint~\eqref{eq:degcons1}, which depends only on the parameter \(\alpha\), vanishes. This is the case for
\begin{equation}
\alpha \in \left\{n\pi + \frac{5\pi}{6}, n \in \mathbb{Z}\right\}\,.
\end{equation}
In this case, the constraint~\eqref{eq:degcons1} is satisfied identically, due to the fact that the two equations~\eqref{eq:degen10} become identical and read
\begin{equation}
-6\dot{H} + 2\dot{K} + (K + 3H)(K - 3H) = 0\,.
\end{equation}
Defining new variables
\begin{equation}
X = 3H + K\,, \quad
Y = 3H - K\,,
\end{equation}
we thus have
\begin{equation}
2\dot{Y} + XY = 0\,,
\end{equation}
and we see that the system is again underdetermined.
\end{enumerate}
We finally remark that also STEGR falls into this class of theories with \(\alpha = 0\). In this case, the constraint~\eqref{eq:degcons1} becomes \(H = 0\), giving Minkowski space as the only spatially flat, homogeneous and isotropic vacuum solution, in agreement with GR, while \(K\) decouples and remains undetermined.

\subsubsection{Branch 2}\label{sssec:degeng2}
We now come to the second branch. If the cosmological parameters take the values~\eqref{eq:degenpar}, there is a unique linear combination of the two dynamical equations~\eqref{eq:dynsys2} so that both \(\dot{H}\) and \(\dot{L}\) cancel from the resulting equation, and one is left with a constraint which reads
\begin{equation}
	\begin{split}
		0 &= \frac{1}{2} \bigg[
		\Big( 3 a_4 K (4H + 3K) + c (5H - 2K + L)(5H + 2K + L) \Big) \cos{\alpha}
		+ \sqrt{3} c \Big( 2K^2 + (H - L)(3H + L) \Big) \sin{3\alpha} \\
		& - c \Big( 13H^2 - 4K^2 + 10HL + L^2 \Big) \cos{3\alpha}
		+ \sqrt{3} \Big(a_4 K (8H + K + 4L) + c (19H^2 + 14HL + 3L^2) \Big) \sin{\alpha} \bigg] \\
		&= \begin{pmatrix}
			H & K & L
		\end{pmatrix} \cdot \mat{M} \cdot \begin{pmatrix}
			H\\
			K\\
			L
		\end{pmatrix}\,,
	\end{split}
\end{equation}
given by the symmetric matrix
\begin{subequations}
	\begin{align}
		M_{11} &=\frac{1}{2}c \Bigg( 25 \cos{\alpha} - 13 \cos{3\alpha} + \sqrt{3} \left( 19 \sin{\alpha} + 3 \sin{3\alpha} \right) \Bigg)  \,,\\
		M_{12} &= a_4 \left( 3 \cos{\alpha} + 2\sqrt{3} \sin{\alpha} \right)\,,\\
		M_{13} &=c \sin{\alpha} \left( 3\sqrt{3} - \sqrt{3} \cos{2\alpha} + 5 \sin{2\alpha} \right) \,,\\
		M_{22} &= \frac{1}{2} \Bigg( (9 a_4 - 4c) \cos{\alpha} + 4c \cos{3\alpha} + \sqrt{3} \left( a_4 \sin{\alpha} + 2c \sin{3\alpha} \right) \Bigg) \,,\\
		M_{23} &= \sqrt{3} a_4 \sin{\alpha} \,,\\
		M_{33} &= 2c \sin^2{\alpha} \left( \cos{\alpha} + \sqrt{3} \sin{\alpha} \right)\,.
	\end{align}
\end{subequations}
The space of solutions to this constraint equation depends on the signature of the appearing bilinear form, expressed by the matrix \(\mat{M}\). One immediately finds that
\begin{equation}
\mat{M} \cdot \begin{pmatrix}
\sin\alpha\\
0\\
-2\sin\alpha - \sqrt{3}\cos\alpha
\end{pmatrix} = 0\,,
\end{equation}
and so \(\mat{M}\) is always degenerate, independently of the values of the parameters \(\alpha, C, a_4\). However, one easily checks that \(\mat{M}\) is non-vanishing unless \(C = a_4 = 0\), which we exclude, since otherwise also the dynamical equations would be trivially solved. In order to simplify the constraint, it is useful to introduce another set \(T, U, V\) of dynamical variables, in terms of which the original variables are given by
\begin{equation}
\begin{pmatrix}
H\\
K\\
L
\end{pmatrix} = \mat{P} \cdot \begin{pmatrix}
T\\
U\\
V
\end{pmatrix}\,,
\end{equation}
where
\begin{equation}
\mat{P} = \begin{pmatrix}
\sin\alpha & 0 & \cos\alpha\\
0 & 1 & 0\\
-2\sin\alpha - \sqrt{3}\cos\alpha & 0 & -2\cos\alpha + \sqrt{3}\sin\alpha
\end{pmatrix}\,.
\end{equation}
In the new variables, the constraint takes the form
\begin{equation}
\begin{split}
0 &= \frac{1}{2}\left[4c\cos(3\alpha) + \sqrt{3}(2c\sin(3\alpha) + a_4\sin\alpha) + (9a_4 - 4c)\cos\alpha\right]U^2 + 6a_4UV + 6c\left(\sqrt{3}\sin\alpha + \cos\alpha\right)V^2\\
&= \begin{pmatrix}
U & V
\end{pmatrix} \cdot \mat{\tilde{M}} \cdot \begin{pmatrix}
U\\
V
\end{pmatrix}\,,
\end{split}
\end{equation}
where the new symmetric matrix \(\mat{\tilde{M}}\) is obtained from the block diagonal form
\begin{equation}
\begin{pmatrix}
0 & 0\\
0 & \mat{\tilde{M}}
\end{pmatrix} = \mat{P}^t \cdot \mat{M} \cdot \mat{P}\,,
\end{equation}
and takes the values
\begin{subequations}
\begin{align}
\tilde{M}_{11} &= \frac{1}{2}\left[4c\cos(3\alpha) + \sqrt{3}(2c\sin(3\alpha) + a_4\sin\alpha) + (9a_4 - 4c)\cos\alpha\right]\,,\\
\tilde{M}_{12} &= 3a_4\,,\\
\tilde{M}_{22} &= 6c\left(\sqrt{3}\sin\alpha + \cos\alpha\right)\,.
\end{align}
\end{subequations}
We thus find that the space of solutions is determined by the signature of the bilinear form defined by \(\mat{\tilde{M}}\). Hence, one has to distinguish the following cases:
\begin{enumerate}
\item
If the signature is \((+, +)\) or \((-, -)\), which is the case for \(\det\mat{\tilde{M}} > 0\), then the solution consists of a single line \(U = V = 0\), and is parametrized by the only free variable \(T\).
\item
If the signature is \((0, +)\) or \((0, -)\), which is the case for \(\det\mat{\tilde{M}} = 0\), then the constraint can be written in the form \((uU + vV)^2 = 0\), with the solution \(U = vS, V = -uS\) in terms of a new variable \(S\), and the solution consists of a single plane spanned by \(S\) and \(T\).
\item
If the signature is \((+, -)\), which is the case for \(\det\mat{\tilde{M}} < 0\), then the constraint can be written in the form \((u_1U + v_1V)(u_2U + v_2V) = 0\), with the two solutions \(U = v_1S, V = -u_1S\) and \(U = v_2S, V = -u_2S\) in terms of a new variable \(S\). Hence, the solution consists of two planes spanned by \(S\) and \(T\), which intersect at \(S = 0\).
\end{enumerate}
Hence, the structure of the solution space is determined by the sign of
\begin{equation}
\det\mat{\tilde{M}} = 3\left\{c\left[-c\cos(4\alpha) + 3(c + a_4)\cos(2\alpha) + \sqrt{3}(6c\cos(2\alpha) - 3c + 5a_4)\sin(2\alpha)\right] - 2c^2 + 6ca_4 - 3a_4^2\right\}\,.
\end{equation}
Studying the full space of solutions to this constraint for general values of the parameters \(\alpha, C, a_4\) is rather tedious and would exceed the scope of this article. Hence, we will restrict ourselves to cases of particular interest in the remainder of this article.

\subsubsection{Branch 3}\label{sssec:degeng3}
We finally consider the third branch. If the cosmological parameters take the values~\eqref{eq:degenpar}, also in this case one finds a unique linear combination of the two dynamical equations~\eqref{eq:dynsys3} so that both \(\dot{H}\) and \(\dot{L}\) cancel from the resulting equation. This yields the constraint
\begin{equation}
	\begin{split}
		0 &=
		\frac{1}{12} \Bigg[
		\sqrt{3} \left( 3 a_4 (4H - K) K + c \left( 9H^2 + 16K^2 + 12KL + L^2 - 6H (2K + L) \right) \right) \cos{\alpha}  \\
		& +\sqrt{3} c \left( 3H^2 - 16K^2 - 12KL - L^2 + 6H (2K + L) \right) \cos{3\alpha} +3 c \left( 3H^2 - 2K^2 - 4KL - L^2 - 2H (6K + L) \right) \sin{3\alpha} \\
		& +3 \left( -a_4 K (3K + 4L) + c \left( 3H^2 + 20K^2 + 12KL + 3L^2 - 2H (6K + L) \right) \right) \sin{\alpha}  \Bigg]
		\\
		&= \begin{pmatrix}
			H & K & L
		\end{pmatrix} \cdot \mat{M} \cdot \begin{pmatrix}
			H\\
			K\\
			L
		\end{pmatrix}\,,
	\end{split}
\end{equation}
which is now given by the symmetric matrix
\begin{subequations}
	\begin{align}
		M_{11} &=c \cos^2{\alpha} \left( \sqrt{3} \cos{\alpha} + 3 \sin{\alpha} \right)  \,,\\
		M_{12} &= \frac{1}{2} \cos{\alpha} \left( \sqrt{3} (a_4 - 2c) + 2\sqrt{3} c \cos{2\alpha} - 6c \sin{2\alpha} \right)\,,\\
		M_{13} &=- c \cos{\alpha} \sin{\alpha} \left( \cos{\alpha} + \sqrt{3} \sin{\alpha} \right) \,,\\
		M_{22} &=- \frac{1}{4} a_4 \left( \sqrt{3} \cos{\alpha} + 3 \sin{\alpha} \right) + \frac{1}{6} c\sin{\alpha} \left( 27 - 6 \cos{2\alpha} + 16 \sqrt{3} \sin{2\alpha} \right) \,,\\
		M_{23} &=- \frac{1}{2} \sin{\alpha} \left( a_4 + 2c \cos{2\alpha} - 2c (1 + \sqrt{3} \sin{2\alpha}) \right)  \,,\\
		M_{33} &=\frac{c\cos{\alpha} \sin^2{\alpha}}{\sqrt{3}} + c \sin^3{\alpha} \,.
	\end{align}
\end{subequations}
As for the second branch, one can determine the space of solutions from the signature of the appearing bilinear form, expressed by the matrix \(\mat{M}\). In this case one finds that
\begin{equation}
\mat{M} \cdot \begin{pmatrix}
\sin\alpha\\
0\\
\sqrt{3}\cos\alpha
\end{pmatrix} = 0\,,
\end{equation}
and so again \(\mat{M}\) is always degenerate, independently of the values of the parameters \(\alpha, C, a_4\). Also here it turns out that \(\mat{M}\) is non-vanishing unless \(C = a_4 = 0\), which we once again exclude as a trivial and hence pathological theory. We can once again simplify the constraint by introducing another set \(T, U, V\) of dynamical variables by defining
\begin{equation}
\begin{pmatrix}
H\\
K\\
L
\end{pmatrix} = \mat{P} \cdot \begin{pmatrix}
T\\
U\\
V
\end{pmatrix}\,,
\end{equation}
where the transformation matrix is now given by
\begin{equation}
\mat{P} = \begin{pmatrix}
\sin\alpha & 0 & \cos\alpha\\
0 & 1 & 0\\
\sqrt{3}\cos\alpha & 0 & -\sqrt{3}\sin\alpha
\end{pmatrix} \cdot \begin{pmatrix}
1 & 0 & 0\\
0 & 2 & 0\\
0 & 2\cos\alpha + 2\sqrt{3}\sin\alpha & 1
\end{pmatrix}\,.
\end{equation}
In the new variables, the constraint takes the form
\begin{equation}
\begin{split}
0 &= \left[(8c + 9a_4)\sin\alpha - 2c\sin(3\alpha) + \frac{8c\cos(3\alpha)}{\sqrt{3}} + \frac{(4c + 9a_4)\cos\alpha}{\sqrt{3}}\right]U^2 + 2\sqrt{3}(2c + a_4)UV + c\left(3\sin\alpha + \sqrt{3}\cos\alpha\right)V^2\\
&= \begin{pmatrix}
U & V
\end{pmatrix} \cdot \mat{\tilde{M}} \cdot \begin{pmatrix}
U\\
V
\end{pmatrix}\,,
\end{split}
\end{equation}
with the symmetric matrix \(\mat{\tilde{M}}\) constructed from the block diagonal form
\begin{equation}
\begin{pmatrix}
0 & 0\\
0 & \mat{\tilde{M}}
\end{pmatrix} = \mat{P}^t \cdot \mat{M} \cdot \mat{P}\,,
\end{equation}
and given by
\begin{subequations}
\begin{align}
\tilde{M}_{11} &= (8c + 9a_4)\sin\alpha - 2c\sin(3\alpha) + \frac{8c\cos(3\alpha)}{\sqrt{3}} + \frac{(4c + 9a_4)\cos\alpha}{\sqrt{3}}\,,\\
\tilde{M}_{12} &= \sqrt{3}(2c + a_4)\,,\\
\tilde{M}_{22} &= c\left(3\sin\alpha + \sqrt{3}\cos\alpha\right)\,.
\end{align}
\end{subequations}
We thus find that the space of solutions is determined by the signature of the bilinear form defined by \(\mat{\tilde{M}}\), which we can study in full analogy to the second branch. In this case the determinant whose sign we must consider is given by
\begin{equation}
\det\mat{\tilde{M}} = c\left[7c\cos(4\alpha) - 9(c + a_4)\cos(2\alpha) + \sqrt{3}(6c\cos(2\alpha) + c + 9a_4)\sin(2\alpha)\right] + 2c^2 + 6ca_4 - 3a_4^2\,.
\end{equation}
As for the second branch, studying the full space of solutions to this constraint for general values of the parameters \(\alpha, C, a_4\) becomes very lengthy, and so we will restrict ourselves to cases of particular interest only.

\subsection{Projective decomposition}\label{ssec:prodecg}
An interesting feature of the dynamical systems derived in section~\ref{ssec:dynsysg} for the non-degenerate case and~\ref{ssec:degeng} for the degenerate case is the fact that the first-order time derivatives of the dynamical variables are expressed as homogeneous second-order polynomials in the dynamical variables. We use this fact in order to perform a decomposition of the dynamics into two components as follows. First, we will denote by \(\vek{z}\) the tuple of dynamical variables, which is given by \(\vek{z} = (H, K)\) for the non-degenerate branch 1, \(\vek{z} = (H, K, L)\) for the non-degenerate branches 2 and 3 and \(\vek{z} = (T)\) or \(\vek{z} = (S, T)\) for the different degenerate cases. Its dynamics then takes the general form
\begin{equation}
\dot{\vek{z}} = \vek{f}(\vek{z}, \vek{z})\,,
\end{equation}
where \(\vek{f}\) denotes a tuple of symmetric bilinear forms. Decomposing \(\vek{z}\) as
\begin{equation}
\vek{z} = Z\vek{n}\,, \quad
Z = \|\vek{z}\|\,, \quad
\vek{n} = \frac{\vek{z}}{\|\vek{z}\|}\,,
\end{equation}
we can write the dynamics as
\begin{equation}
\dot{\vek{z}} = \dot{Z}\vek{n} + Z\dot{\vek{n}} = Z^2\vek{f}(\vek{n}, \vek{n})\,,
\end{equation}
where
\begin{subequations}
\begin{align}
\dot{Z} &= \dot{\vek{z}} \cdot \vek{n} = Z^2\vek{f}(\vek{n}, \vek{n}) \cdot \vek{n}\,,\\
\dot{\vek{n}} &= \frac{\dot{\vek{z}} - (\dot{\vek{z}} \cdot \vek{n})\vek{n}}{Z} = Z\left\{\vek{f}(\vek{n}, \vek{n}) - \left[\vek{f}(\vek{n}, \vek{n}) \cdot \vek{n}\right]\vek{n}\right\}\,.
\end{align}
\end{subequations}
This decomposition is valid whenever \(\|\vek{z}\| \neq 0\), and so we have to restrict the dynamics accordingly; however, this is not a significant restriction, since \(\vek{z} = 0\) is always a fixed point, as detailed in the following section, and so it can be treated separately. We will henceforth assume \(Z > 0\). It then follows that the qualitative dynamics of the system, given by the sign of \(\dot{Z}\) and the direction of \(\dot{\vek{n}}\), is fully determined by \(\vek{n}\) only. We will make use of this fact in the following sections in order to determine the qualitative properties of the dynamics.

\subsection{Fixed points}\label{ssec:fixpg}
An important property of any dynamical system are fixed points, i.e., solutions of the equation
\begin{equation}
\dot{\vek{z}} = \vek{f}(\vek{z}, \vek{z}) = 0\,.
\end{equation}
We will denote such solutions as \(\vek{z}^*\). It follows immediately from the discussion of the previous section that for all branches of dynamical systems we derived the origin \(\vek{z}^* = 0\) of the dynamical phase space is always a fixed point, since
\begin{equation}
\vek{f}(0, 0) = 0
\end{equation}
due to the homogeneity of \(\vek{f}\). To assess the stability of this fixed point, which follows from the dynamics of the system in its vicinity, we again use the homogeneity, from which further follows
\begin{equation}
\vek{f}(\vek{z}, \vek{z}) = \vek{f}(-\vek{z}, -\vek{z})\,.
\end{equation}
Hence, for every vector \(\vek{z} = Z\vek{n}\) there exists a vector \(\vek{z}' = Z'\vek{n}' = -\vek{z}\), with \(Z' = Z\) and \(\vek{n}' = -\vek{n}\), such that \(\dot{\vek{z}}' = \dot{\vek{z}}\), and therefore \(\dot{Z}' = -\dot{Z}\) and \(\dot{\vek{n}}' = \dot{\vek{n}}\). From the former in particular follows that if the trajectories are approaching the origin \(\vek{z} = 0\), and hence \(\dot{Z} < 0\), then \(\dot{Z}' > 0\), and so the trajectories diverge on the opposite side. This fixed point is therefore always a (non-hyperbolic) saddle point.

For generic values of the cosmological parameters, the origin \(\vek{z}^* = 0\) is the only fixed point of all dynamical systems we study here. For particular values of the parameters, however, further fixed points appear. Note that the fixed point condition
\begin{equation}
0 = \dot{\vek{z}} = \vek{f}(\vek{z}, \vek{z}) = Z^2\vek{f}(\vek{n}, \vek{n})
\end{equation}
depends only on \(\vek{n}\) if \(Z > 0\), and so if \(\vek{z}^*\) is a fixed point, the same is true for any scalar multiple \(c\vek{z}^*\) of \(\vek{z}^*\) with \(c \in \mathbb{R}\). This has an important consequence for the stability of these fixed points, which we can see as follows. Recall that the stability of a fixed point can generically be determined by the eigenvalues of the Jacobi matrix \(\mat{J}\), whose value \(\mat{J}^*\) at a fixed point \(\vek{z}^*\) is defined by a linear approximation \(\vek{z} = \vek{z}^* + \delta\vek{z}\) as
\begin{equation}
\dot{\vek{z}} = \vek{f}(\vek{z}^* + \delta\vek{z}, \vek{z}^* + \delta\vek{z}) = \mat{J}^* \cdot \delta\vek{z} + O(\delta\vek{z}^2)\,.
\end{equation}
If \(\delta\vek{z} = c\vek{z}^*\) is a scalar multiple of \(\vek{z}^*\), then it follows that
\begin{equation}
\dot{\vek{z}} = \vek{f}(\vek{z}^* + c\vek{z}^*, \vek{z}^* + c\vek{z}^*) = (c + 1)^2\vek{f}(\vek{z}^*, \vek{z}^*) = 0\,,
\end{equation}
and so the Jacobi matrix \(\mat{J}^*\) always has at least one vanishing eigenvalue. Further, for a scalar multiple \(\vek{z}'^{\star} = c\vek{z}^{\star}\) with \(c \neq 0\) we have
\begin{equation}
\begin{split}
\mat{J}'^* \cdot \delta\vek{z} + O(\delta\vek{z}^2) &= \dot{\vek{z}}'\\
&= \vek{f}(c\vek{z}^* + \delta\vek{z}, c\vek{z}^* + \delta\vek{z})\\
&= c^2\vek{f}(\vek{z}^* + c^{-1}\delta\vek{z}, \vek{z}^* + c^{-1}\delta\vek{z})\\
&= c^2\mat{J}^* \cdot c^{-1}\delta\vek{z} + O(\delta\vek{z}^2)\\
&= c\mat{J}^* \cdot \delta\vek{z} + O(\delta\vek{z}^2)\,,
\end{split}
\end{equation}
and so \(\mat{J}'^* = c\mat{J}^*\). In particular, for \(c < 0\), the eigenvalues of \(\mat{J}'^*\) will have the opposite sign compared to the eigenvalues of \(\mat{J}^*\), and so their stability will be reversed. The relevance of this observation will become clear when we discuss the fixed points of particular theories.

\subsection{Projective fixed points}\label{ssec:profixg}
In section~\ref{ssec:prodecg} we have seen that the dynamics of the flat newer GR vacuum cosmology we study in this article can be decomposed into angular and radial parts. We now come to an interesting consequence of this decomposition. For this purpose, note that if \(\vek{n}\) satisfies the condition
\begin{equation}\label{eq:profixcond}
\frac{\dot{\vek{n}}}{Z} = \vek{f}(\vek{n}, \vek{n}) - \left[\vek{f}(\vek{n}, \vek{n}) \cdot \vek{n}\right]\vek{n} = 0\,,
\end{equation}
then also \(\dot{\vek{n}} = 0\), i.e., \(\vek{n}\) is constant. We will denote points \(\vek{z}\) of the phase space at which this condition holds by \(\vek{z}^{\star}\) and call their normalized vectors \(\vek{n}^{\star} = \vek{z}^{\star}/\|\vek{z}^{\star}\|\) \emph{projective fixed points}.

Since the angular dynamics depends only on \(\vek{n}\) up to a positive (but generally time-dependent) factor \(Z\), it follows that the stability of these projective fixed points, which is determined by the direction of \(\dot{\vek{n}}\), also depends on \(\vek{n}\) only. Similarly to the case of fixed points discussed in the previous section, their stability can thus be determined by the eigenvalues of the Jacobi matrix \(\mat{J}^{\star}\), which we now define by an expansion \(\vek{z} = \vek{z}^{\star} + \delta\vek{z}\) around a projective fixed point \(\vek{z}^{\star}\), so that
\begin{equation}
\frac{\dot{\vek{n}}}{Z} = \frac{\vek{f}(\vek{z}^{\star} + \delta\vek{z}, \vek{z}^{\star} + \delta\vek{z})}{\|\vek{z}^{\star} + \delta\vek{z}\|^2} - \frac{\left[\vek{f}(\vek{z}^{\star} + \delta\vek{z}, \vek{z}^{\star} + \delta\vek{z}) \cdot (\vek{z}^{\star} + \delta\vek{z})\right](\vek{z}^{\star} + \delta\vek{z})}{\|\vek{z}^{\star} + \delta\vek{z}\|^4} = \mat{J}^{\star} \cdot \delta\vek{z} + O(\delta\vek{z}^2)\,.
\end{equation}
By explicit calculation one immediately finds that
\begin{equation}
\mat{J}^{\star} \cdot \vek{z}^{\star} = 0\,,
\end{equation}
and so the Jacobi matrix always has a vanishing eigenvalue, whose eigenvector is pointing in the radial direction, which also follows from the fact that \(\dot{\vek{n}}/Z\) does not depend on \(Z\).

Further, for these points also
\begin{equation}
N^{\star} = \vek{f}(\vek{n}^{\star}, \vek{n}^{\star}) \cdot \vek{n}^{\star}
\end{equation}
is constant, and so the radial dynamics takes the simple form
\begin{equation}
\dot{Z} = N^{\star}Z^2\,,
\end{equation}
which can be solved explicitly and yields the general solution
\begin{equation}
Z(t) = \frac{1}{N^{\star}(t_0 - t)}\,,
\end{equation}
for \(N^{\star} \neq 0\), where \(t_0\) is a constant of integration, and we must impose \(t < t_0\) for \(N^{\star} > 0\) leading to a future finite time singularity at \(t_0\), and \(t > t_0\) for \(N^{\star} < 0\) leading to a past finite time singularity at \(t_0\), while \(Z(t) = \text{const.}\) for \(N^{\star} = 0\).

It further follows from the reflection symmetry of the condition~\eqref{eq:profixcond} that for each projective fixed point \(\vek{n} = \vek{n}^{\star}\) also \(-\vek{n} = \vek{n}'^{\star}\) is a projective fixed point, and that \(\mat{J}'^{\star} = -\mat{J}^{\star}\) and \(N'^{\star} = -N^{\star}\). This means that antipodal projective fixed points have opposite stability properties, i.e., if one of them is an attractor, then the other one is a repeller, and vice versa. Further, if one of them leads to a future finite time singularity, then the other one leads to a past finite time singularity. Due to this particular pairwise appearance of projective fixed points, it is not sufficient to study the question of their stability in isolation, but one needs to consider its correlation with the radial dynamics, in order to determine whether there is an attractive projective fixed point pulling trajectories towards a future finite time singularity, or towards the central fixed point \(\vek{z} \to 0\). Further, one also needs to consider the correlation of these properties with the sign of \(H\), in order to determine which of these scenarios occurs in an expanding or a collapsing universe. In the following sections of this article, we will study the existence and properties of projective fixed points for the different branches of cosmological solutions.

\subsection{Effective dark energy}\label{ssec:effdarkeng}
In order to compare the dynamics of the discussed class of teleparallel gravity models to that of the $\Lambda$CDM model, we rewrite the cosmological vacuum equations in the form of modified Friedmann equations as
\begin{equation}
3H^2 = \kappa^2\rho_{\lambda}\,, \quad
-2\dot{H} - 3H^2 = \kappa^2p_{\lambda}\,,
\end{equation}
where we introduced the effective energy density \(\rho_{\lambda}\) and pressure \(p_{\lambda}\) of dark energy. The effective barotropic index is thus given by
\begin{equation}\label{eq:baroin}
w_{\lambda} = \frac{p_{\lambda}}{\rho_{\lambda}} = -1 - \frac{2\dot{H}}{3H^2}\,.
\end{equation}
We derive the effective barotropic index for different types of theories and cosmological branches in the following sections.

\section{Theories of type 1}\label{sec:type1}
After the general discussion in the previous section, we now turn our focus to the theories of type 1, for which the parameters in the action take the form~\eqref{eq:ngrtype1}, so that the parameters~\eqref{eq:cosmopar} in the cosmological field equations become
\begin{equation}\label{eq:partype1}
a_1 = 1\,, \quad
a_2 = 0\,, \quad
a_3 = 0\,, \quad
a_4 = \epsilon\,.
\end{equation}
Following the derivation detailed in section~\ref{ssec:degeng}, we thus see that these theories lead to a degenerate linear system, which corresponds to the parametrization~\eqref{eq:degenpar} with \(C = 1\) and \(\alpha = 0\). We have seen in section~\ref{sssec:degeng1} that for the first branch of cosmological solutions, this system reduces to the dynamics of STEGR. We can therefore restrict ourselves to a study of the remaining two branches, and we will assume \(\epsilon \neq 0\) to exclude the STEGR case. We discuss the second branch in section~\ref{ssec:type12} and the third branch in section~\ref{ssec:type13}.

\subsection{Branch 2}\label{ssec:type12}
We start with the second branch. For the values~\eqref{eq:partype1} of the cosmological parameters, equation~\eqref{eq:dynsys21} becomes a constraint, which reads
\begin{equation}
3H^2 + 3\epsilon HK + \frac{9}{4}\epsilon K^2 = 0\,.
\end{equation}
This constraint has real solutions only for \(\epsilon \leq 0\) or \(\epsilon \geq 3\). With the previously mentioned restriction \(\epsilon \neq 0\), we can solve for \(K\) and obtain
\begin{equation}
K = -\frac{2}{3}\left(1 \pm \sqrt{1 - \frac{3}{\epsilon}}\right)H\,.
\end{equation}
Further, we find that \(\dot{L}\) vanishes from the dynamical equations, and we can eliminate \(L\) by solving the condition
\begin{equation}
\dot{H} = HL\,,
\end{equation}
which follows from the definition~\eqref{eq:vardef} together with the solution for \(K\) given above and its time derivative. Substituting these results in the final equation~\eqref{eq:dynsys22}, which reads
\begin{equation}
-2\dot{H} - 3H^2 + \epsilon K\left(2H + \frac{K}{4} + L\right) = 0\,,
\end{equation}
one needs to distinguish two cases. If \(\epsilon = -1\) and
\begin{equation}
K = -\frac{2}{3}\left(1 + \sqrt{1 - \frac{3}{\epsilon}}\right)H = -2H\,,
\end{equation}
we find that the final equation is satisfied identically and the dynamics of \(H\) is undetermined. If the other solution for \(K\) is chosen, or \(\epsilon \neq -1\), one obtains the equation
\begin{equation}
3\dot{H} + 5H^2 = 0\,,
\end{equation}
which is now independent of \(\epsilon\). The general solution to this equation is given by
\begin{equation}
H(t) = \frac{3}{5(t - t_0)}\,,
\end{equation}
where \(t_0\) is an integration constant. This solution describes a decelerating expansion for \(t > t_0\) with a constant effective barotropic index of \(w_{\lambda} = 1/9\).

\subsection{Branch 3}\label{ssec:type13}
The analysis of the third branch follows the same steps. Inserting the values~\eqref{eq:partype1}, equation~\eqref{eq:dynsys31} becomes a constraint and reads
\begin{equation}
3H^2 + 3\epsilon HK - \frac{3}{4}\epsilon K^2 = 0\,.
\end{equation}
In this case we find that real solutions exist for \(\epsilon \geq 0\) or \(\epsilon \leq -1\). Assuming \(\epsilon \neq 0\), we can solve for \(K\) and find
\begin{equation}
K = 2\left(1 \pm \sqrt{1 + \frac{1}{\epsilon}}\right)H\,.
\end{equation}
Also in this case we find that \(\dot{L}\) vanishes from the dynamical equations, and we can eliminate \(L\) by solving the same condition
\begin{equation}
\dot{H} = HL\,,
\end{equation}
which again follows from the definition~\eqref{eq:vardef} together with the solution for \(K\) given above and its time derivative. Now substituting these results in the final equation~\eqref{eq:dynsys32}, which in this case reads
\begin{equation}
-2\dot{H} - 3H^2  \epsilon K\left(\frac{3}{4}K + L\right) = 0\,,
\end{equation}
one again needs to distinguish two cases. If \(\epsilon = -1\) and thus
\begin{equation}
K = 2\left(1 \pm \sqrt{1 + \frac{1}{\epsilon}}\right)H = 2H\,,
\end{equation}
we find again that the final equation is satisfied identically and the dynamics of \(H\) is undetermined. If \(\epsilon \neq -1\), one obtains the equation
\begin{equation}
\dot{H} + 3H^2 = 0\,,
\end{equation}
which is now independent of \(\epsilon\). The general solution to this equation is given by
\begin{equation}
H(t) = \frac{1}{3(t - t_0)}\,,
\end{equation}
where \(t_0\) is an integration constant. This solution describes a decelerating expansion for \(t > t_0\) with a constant effective barotropic index of \(w_{\lambda} = 1\).

\section{Theories of type 2}\label{sec:type2}
We now turn our focus to the theories of type 2, in which the parameters in the action take the values~\eqref{eq:ngrtype2}. In this case the cosmological parameters~\eqref{eq:cosmopar} become
\begin{equation}\label{eq:partype2}
a_1 = 1 + 3\epsilon\,, \quad
a_2 = 2\epsilon\,, \quad
a_3 = \epsilon\,, \quad
a_4 = 0\,.
\end{equation}
From these parameter values further follows
\begin{equation}
4a_1a_3 - 3a_2^2 = 4\epsilon\,,
\end{equation}
and so by comparison with the condition~\eqref{eq:degencond} derived in section~\ref{ssec:degeng} we find that these theories are non-degenerate unless \(\epsilon = 0\), which we exclude, since otherwise the theory would reduce to STEGR. We can thus apply the dynamical system formulation constructed in section~\ref{ssec:dynsysg}, which we will do in section~\ref{ssec:dynsys2}. We then discuss different properties of this dynamical system. In particular, we study effective dark energy in section~\ref{ssec:effdarken2}, bounces and turnarounds in section~\ref{ssec:bt2}, fixed points in section~\ref{ssec:fixp2}, projective fixed points in section~\ref{ssec:profix2} and the complete phase space in section~\ref{ssec:phase2}.

\subsection{Dynamical system}\label{ssec:dynsys2}
Since we only consider the non-degenerate case \(\epsilon \neq 0\), it follows that we can use the dynamical system formulation derived in detail in section~\ref{ssec:dynsysg}. We then find that for the first branch, the dynamical system~\eqref{eq:nondeg1} becomes
\begin{subequations}
\begin{align}
\dot{H} &= \epsilon(3H + K)^2\,,\\
\dot{K} &= -\frac{1}{2}(3H + K)(9H + K) - 3\epsilon(3H + K)^2 - \frac{3H^2}{2\epsilon}\,,
\end{align}
\end{subequations}
while we find
\begin{subequations}
\begin{align}
\dot{H} &= \epsilon(5H + L)^2\,,\\
\dot{K} &= KL\,,\\
\dot{L} &= -\frac{1}{2}(5H + L)(11H + L) - 5\epsilon(5H + L)^2 - \frac{3H^2}{2\epsilon}\,,
\end{align}
\end{subequations}
for the second branch~\eqref{eq:nondeg2}, and finally
\begin{subequations}
\begin{align}
\dot{H} &= \epsilon(3H - 4K - L)^2\,,\\
\dot{K} &= KL\,,\\
\dot{L} &= \frac{27}{2}H^2 - 6H(4K + L) + 8K^2 + \frac{1}{2}L^2 + 3\epsilon(3H - 4K - L)^2 + \frac{3H^2}{2\epsilon}\,,
\end{align}
\end{subequations}
for the third branch~\eqref{eq:nondeg3}.

\subsection{Effective dark energy}\label{ssec:effdarken2}
Using the expressions for \(\dot{H}\) for the dynamical system found in the previous section in the formula~\eqref{eq:baroin}, we find that the effective barotropic index to be
\begin{equation}\label{eq:baroin1}
w_{\lambda} = -1 - 2\epsilon\frac{(3H + K)^2}{3H^2}
\end{equation}
in the first branch,
\begin{equation}\label{eq:baroin2}
w_{\lambda} = -1 - 2\epsilon\frac{(5H + L)^2}{3H^2}
\end{equation}
in the second branch and
\begin{equation}\label{eq:baroin3}
w_{\lambda} = -1 - 2\epsilon\frac{(3H - 4K - L)^2}{3H^2}
\end{equation}
in the third branch. From these expressions we see that for \(\epsilon > 0\) the barotropic index satisfies \(w_{\lambda} < -1\) almost everywhere in the phase space of the dynamical system, with the only exception given by a single hyperplane, where the numerator given in the corresponding equation above vanishes, and so the barotropic index becomes \(w_{\lambda} = -1\), and the hyperplane \(H = 0\), where it is singular. Hence, in these models the effective contribution of modified gravity to the cosmological field equations behaves as phantom dark energy, except for a hyperplane in the phase space, where it becomes vacuum energy. For \(\epsilon < 0\), the opposite behavior is observed, and we find \(w_{\lambda} > -1\) almost everywhere, so no phantom dark energy appears.

\subsection{Bounce and turnaround}\label{ssec:bt2}
We now study the possibility of bounces and turnarounds for the different cosmological solution branches, which can be done by studying \(\dot{H}\) at \(H = 0\). A turnaround occurs if \(\dot{H} < 0\), while a bounce has \(\dot{H} > 0\). From the dynamical equations derived in section~\ref{ssec:dynsys2}, we see that for \(\epsilon > 0\) we always have \(\dot{H} \geq 0\), and so only bounces are possible, while turnarounds are excluded. The opposite is the case for \(\epsilon < 0\), where \(\dot{H} \leq 0\).

\subsection{Fixed points}\label{ssec:fixp2}
Comparing the fixed point conditions detailed in section~\ref{ssec:fixpg} with the parameter values~\eqref{eq:partype2}, we find that the only condition which is satisfied is the condition \(a_2 = 2a_3 - a_4\), which means that in the second branch the whole line \(H = L = 0\) consists of fixed points, which turn out to be saddle points. This result is independent of the parameter \(\epsilon\). For the other two branches, only the trivial fixed points \(H = K = 0\) in the first branch and \(H = K = L = 0\) in the second and third branch exist.

\subsection{Projective fixed points}\label{ssec:profix2}
One of the most interesting features of theories of type 2 is the existence of numerous projective fixed points, as we discussed in the general case in section~\ref{ssec:profixg}. In the following, we construct the complete list of projective fixed points, along with their stability and radial dynamics, depending on the value of the parameter \(\epsilon\).

\subsubsection{Branch 1}\label{sssec:profixp21}
The projective fixed point condition~\eqref{eq:profixcond} for the first branch reads
\begin{equation}\label{eq:profixcond21}
\frac{(H + 6\epsilon H + 2\epsilon K)\left[3H^2 + \epsilon(3H + K)^2\right]}{2\epsilon} = 0\,.
\end{equation}
Keeping in mind the assumption \(\epsilon \neq 0\) to exclude the STEGR case, we thus find the following solutions:

\begin{enumerate}
\item
From the linear factor in the condition~\eqref{eq:profixcond21} we find the solution
\begin{equation}\label{eq:profix211}
K = -\frac{1 + 6\epsilon}{2\epsilon}H\,.
\end{equation}
For this projective fixed point, the Jacobi matrix is given by
\begin{equation}
Z^2\mat{J}^{\star} = -\frac{\epsilon(1 + 12\epsilon)}{4\epsilon^2(1 + 12\epsilon + 40\epsilon^2)}H\begin{pmatrix}
(1 + 6\epsilon)^2 & 2\epsilon(1 + 6\epsilon)\\
2\epsilon(1 + 6\epsilon) & 4\epsilon^2
\end{pmatrix}
\end{equation}
and has the single non-trivial eigenvalue
\begin{equation}
Z^2\lambda^{\star} = -\left(3 + \frac{1}{4\epsilon}\right)H\,.
\end{equation}
Hence, we find that for \(\epsilon > 0\) and \(\epsilon < -1/12\) this is an attractor for \(H > 0\) and a repeller for \(H < 0\), while for \(-1/12 < \epsilon < 0\) the opposite is the case. For \(\epsilon = -1/12\), it is a non-hyperbolic projective fixed point. Finally, we find that the radial dynamics is given by
\begin{equation}
\dot{\vek{z}} = \frac{H}{4\epsilon}\vek{z}\,.
\end{equation}
It follows that for \(\epsilon > 0\) this projective fixed point yields a future finite time singularity for \(H > 0\) and a past finite time singularity for \(H < 0\), while for \(\epsilon < 0\) the opposite occurs. Finally, for the effective barotropic index~\eqref{eq:baroin1} at this point we find
\begin{equation}
w_{\lambda} = -1 - \frac{1}{6\epsilon}\,,
\end{equation}
and we will discuss the relevance of this result when we study the phase space in section~\ref{sssec:phase21}.

\item
For \(\epsilon < 0\), the quadratic factor in the condition~\eqref{eq:profixcond21} yields another pair of solutions given by
\begin{equation}\label{eq:profix212}
K = \left(-3 \pm \sqrt{-\frac{3}{\epsilon}}\right)H\,.
\end{equation}
In this case, we find that the Jacobi matrix is given by
\begin{equation}
Z^2\mat{J}^{\star} = -\frac{6\epsilon + \sqrt{-3\epsilon}}{\epsilon\left(10\epsilon + 6\sqrt{-3\epsilon} - 3\right)}H\begin{pmatrix}
\left(3\sqrt{-\epsilon} - \sqrt{3}\right)^2 & \pm\sqrt{-3\epsilon}\left(\sqrt{-3\epsilon} - 1\right)\\
\pm\sqrt{-3\epsilon}\left(\sqrt{-3\epsilon} - 1\right) & -\epsilon
\end{pmatrix}
\end{equation}
and has the non-trivial eigenvalue
\begin{equation}
Z^2\lambda^{\star} = \left(6 \mp \sqrt{-\frac{3}{\epsilon}}\right)H\,.
\end{equation}
This shows that the two fixed points, which are distinguished by the upper and lower sign, have different stability properties. For the lower sign, we find a repeller for \(H > 0\) and an attractor for \(H < 0\). For the upper sign, the stability depends on the value of \(\epsilon\). If \(\epsilon < -1/12\), the behavior is identical to the previously discussed case of the lower sign, while for \(-1/12 < \epsilon < 0\) we find the opposite behavior. For \(\epsilon = -1/12\), we find a non-hyperbolic projective fixed point. For the radial dynamics, we finally find
\begin{equation}
\dot{\vek{z}} = -3H\vek{z}
\end{equation}
for both fixed points, independently of the parameter \(\epsilon\). Hence, for both points \(H > 0\) corresponds to a past finite time singularity, while \(H < 0\) yields a future finite time singularity. For both fixed points we find the barotropic index~\eqref{eq:baroin1} becomes \(w_{\lambda} = 1\), and hence the modified gravity contribution to the cosmological dynamics behaves as a stiff fluid.
\end{enumerate}

\subsubsection{Branch 2}\label{sssec:profixp22}
Next, we consider the second branch. In this case the condition~\eqref{eq:profixcond} reads
\begin{subequations}\label{eq:profixcond22}
\begin{align}
\frac{1}{2}K\left[\frac{3H^2}{\epsilon} + 55H^2 + 10\epsilon(5H + L)^2 + 16HL + 3L^2\right] &= 0\,,\\
-\frac{(H + 10\epsilon H + 2\epsilon L) + H)\left[3H^2 + \epsilon(5H + L)^2\right]}{2\epsilon} &= 0\,,\\
K\left[\epsilon(5H + L)^2 - HL\right] &= 0\,.
\end{align}
\end{subequations}
This system has the following solutions.

\begin{enumerate}
\item
Note first that
\begin{equation}\label{eq:profix221}
H = L = 0
\end{equation}
is obviously a projective fixed point. Following the same procedure as for the first branch, one finds that the Jacobi matrix vanishes identically, and so this is a non-hyperbolic projective fixed point. Calculating the radial dynamics, one finds
\begin{equation}
\dot{\vek{z}} = 0\,,
\end{equation}
and so this line, in fact, corresponds to a line of fixed points, in agreement to our finding from section~\ref{ssec:fixp2}. Note that the barotropic index~\eqref{eq:baroin2} is undefined on this line, and takes different values in its neighborhood.

\item
Another projective fixed point is located at
\begin{equation}\label{eq:profix222}
K = 0\,, \quad
L = -\left(5 + \frac{1}{2\epsilon}\right)H\,,
\end{equation}
for which we now find the Jacobi matrix
\begin{equation}
Z^2\mat{J}^{\star} = -\frac{\epsilon(12\epsilon + 1)}{4\epsilon^2(104\epsilon^2 + 20\epsilon + 1)}H\begin{pmatrix}
(1 + 10\epsilon)^2 & 0 & 2\epsilon(1 + 10\epsilon)\\
0 & \frac{(3 + 20\epsilon)(1 + 20\epsilon + 104\epsilon^2)}{1 + 12\epsilon} & 0\\
2\epsilon(1 + 10\epsilon) & 0 & 4\epsilon^2
\end{pmatrix}\,.
\end{equation}
One finds that its eigenvalues are given by
\begin{equation}
Z^2\lambda^{\star}_1 = -\left(3 + \frac{1}{4\epsilon}\right)H\,, \quad
Z^2\lambda^{\star}_2 = -\left(5 + \frac{3}{4\epsilon}\right)H\,,
\end{equation}
from which the stability of this projective fixed point can be derived as follows. For \(\epsilon > 0\) or \(\epsilon < -\frac{3}{20}\), we find that this fixed point is an attractor for \(H > 0\) and a repeller for \(H < 0\). At \(\epsilon = -\frac{3}{20}\), we have \(\lambda^{\star}_2 = 0\), so that it becomes non-hyperbolic. In the range \(-\frac{3}{20} < \epsilon < -\frac{1}{12}\), the two eigenvalues have opposite signs, and so it becomes a saddle point. Then at \(\epsilon = -\frac{1}{12}\), we find \(\lambda^{\star}_1 = 0\), and it again becomes non-hyperbolic. Finally, for \(-\frac{1}{12} < \epsilon < 0\), we find that this point is an attractor for \(H < 0\) and a repeller for \(H > 0\). For the radial dynamics, we obtain the relation
\begin{equation}
\dot{\vek{z}} = \frac{H}{4\epsilon}\vek{z}\,,
\end{equation}
which shows that for \(\epsilon > 0\) one has a future finite time singularity for \(H > 0\) and a past finite time singularity for \(H < 0\), while for \(\epsilon < 0\) the opposite is the case. Here we find the barotropic index~\eqref{eq:baroin2}
\begin{equation}
w_{\lambda} = -1 - \frac{1}{6\epsilon}\,,
\end{equation}
similar to the first branch.

\item
If \(\epsilon < 0\), another pair of projective fixed points exists at
\begin{equation}\label{eq:profix223}
K = 0\,, \quad
L = \left(-5 \pm \sqrt{-\frac{3}{\epsilon}}\right)H\,.
\end{equation}
In this case the Jacobi matrix takes the form
\begin{equation}
Z^2\mat{J}^{\star} = \frac{\sqrt{-\epsilon}\left(6\sqrt{-\epsilon} \mp \sqrt{3}\right)}{\epsilon\left(26\epsilon \pm 10\sqrt{-3\epsilon} - 3\right)}H\begin{pmatrix}
\left(5\sqrt{-\epsilon} \mp \sqrt{3}\right)^2 & 0 & \sqrt{-\epsilon}\left(5\sqrt{-\epsilon} \mp \sqrt{3}\right)\\
0 & \frac{\left(2\sqrt{-\epsilon} \mp \sqrt{3}\right)\left(26\epsilon \pm 10\sqrt{-3\epsilon} - 3\right)}{6\sqrt{-\epsilon} \mp \sqrt{3}} & 0\\
\sqrt{-\epsilon}\left(5\sqrt{-\epsilon} \mp \sqrt{3}\right) & 0 & -\epsilon
\end{pmatrix}
\end{equation}
and possesses the eigenvalues
\begin{equation}
Z^2\lambda^{\star}_1 = \left(-2 \pm \sqrt{-\frac{3}{\epsilon}}\right)H\,, \quad
Z^2\lambda^{\star}_2 = \left(6 \mp \sqrt{-\frac{3}{\epsilon}}\right)H\,.
\end{equation}
Hence, the stability of these projective fixed points depends on \(\epsilon\) as follows. First, note that the projective fixed point denoted by the lower sign is always a saddle point, since in this case the two eigenvalues always have opposite signs, independently of \(\epsilon\). The same holds true for the other projective fixed point given by the upper sign if \(\epsilon < -\frac{3}{4}\) or \(-\frac{1}{12} < \epsilon < 0\). At \(\epsilon = -\frac{3}{4}\), we have \(\lambda^{\star}_1 = 0\), while \(\lambda^{\star}_2 = 0\) at \(\epsilon = -\frac{1}{12}\), and so this fixed point becomes non-hyperbolic. Finally, for \(-\frac{3}{4} < \epsilon < -\frac{1}{12}\), we find a repeller for \(H > 0\) and an attractor for \(H < 0\). Taking a look at the radial dynamics, we find
\begin{equation}
\dot{\vek{z}} = -3H\vek{z}\,,
\end{equation}
which shows that both projective fixed points correspond to past finite time singularities for \(H > 0\) and future finite time singularities for \(H < 0\). The barotropic index~\eqref{eq:baroin2} takes the stiff fluid value \(w_{\lambda} = 1\).
\end{enumerate}

\subsubsection{Branch 3}\label{sssec:profixp23}
Finally, we consider also the third branch. The projective fixed point condition~\eqref{eq:profixcond} now becomes
\begin{subequations}\label{eq:profixcond23}
\begin{align}
\frac{K}{2 \epsilon}\left\{\epsilon\left[-27H^2 + 12H(4K + L) - 16K^2 + L^2\right] - 3H^2 - 6\epsilon^2(4K + L - 3H)^2\right\} &= 0\,,\\
\frac{3 H^3}{2\epsilon} + \frac{H}{2}\left[27H^2 - 12H(4K + L) + 16K^2 + L^2\right] + \epsilon(3H - L)(4K + L - 3H)^2 &= 0\,,\\
K\left[\epsilon(4K + L - 3H)^2 - HL\right] &= 0\,.
\end{align}
\end{subequations}
The solutions are given as follows.

\begin{enumerate}
\item
Independently of \(\epsilon\), one always finds the projective fixed point
\begin{equation}\label{eq:profix231}
H = 0\,, \quad
L = -4K\,,
\end{equation}
for which the Jacobi matrix becomes
\begin{equation}
Z^2\mat{J}^{\star} = \frac{4}{17}K\begin{pmatrix}
17 & 0 & 0\\
0 & 16 & 4\\
0 & 4 & 1
\end{pmatrix}
\end{equation}
and possesses two identical non-trivial eigenvalues
\begin{equation}
Z^2\lambda_{1,2}^{\star} = 4K\,.
\end{equation}
It follows that this projective fixed point is a repeller for \(K > 0\) and an attractor for \(K < 0\). The radial dynamics is governed by the relation
\begin{equation}
\dot{\vek{z}} = -4K\vek{z}\,,
\end{equation}
which shows that it is a past finite time singularity for \(K > 0\) and a future finite time singularity for \(K < 0\). Finally, we find that the barotropic index~\eqref{eq:baroin3} is undefined.

\item
Another projective fixed point is located at
\begin{equation}\label{eq:profix232}
K = 0\,, \quad
L = \left(3 + \frac{1}{2\epsilon}\right)H\,.
\end{equation}
In this case one finds the Jacobi matrix
\begin{equation}
Z^2\mat{J}^{\star} = \frac{12\epsilon + 1}{4\epsilon^2(40\epsilon^2 + 12\epsilon + 1)}H\begin{pmatrix}
-(6\epsilon + 1)^2 & 16\epsilon(6\epsilon + 1) & 2\epsilon(6\epsilon + 1)\\
0 & 40\epsilon^2 + 12\epsilon + 1 & 0\\
2\epsilon(6\epsilon + 1) & -32\epsilon^2 & -4\epsilon^2
\end{pmatrix}\,,
\end{equation}
together with the eigenvalues
\begin{equation}
Z^2\lambda_{1,2}^{\star} = \pm\left(3 + \frac{1}{4\epsilon}\right)H\,.
\end{equation}
Note that these eigenvalues always have opposite signs, which means that this point is always a saddle point, independently of \(\epsilon\), which becomes non-hyperbolic at \(\epsilon = -\frac{1}{12}\). For the radial dynamics, one finds the relation
\begin{equation}
\dot{\vek{z}} = \frac{H}{4\epsilon}\vek{z}\,.
\end{equation}
Hence, for \(\epsilon > 0\) one obtains a future finite time singularity for \(H > 0\) and a past finite time singularity for \(H < 0\), while the opposite occurs for \(\epsilon < 0\). In this case the barotropic index~\eqref{eq:baroin3} takes the value
\begin{equation}
w_{\lambda} = -1 - \frac{1}{6\epsilon}\,.
\end{equation}

\item
Next, we consider the projective fixed point
\begin{equation}\label{eq:profix233}
K = \left(\frac{3}{4} + \frac{1}{16\epsilon}\right)H\,, \quad
L = \frac{H}{4\epsilon}\,,
\end{equation}
for which we find the Jacobi matrix
\begin{equation}
Z^2\mat{J}^{\star} = -\frac{12\epsilon + 1}{4\epsilon(400\epsilon^2 + 24\epsilon + 17)}H\begin{pmatrix}
24\epsilon(6\epsilon + 1) + 17 & -16\epsilon(12\epsilon + 1) & -64\epsilon\\
-16\epsilon(12\epsilon + 1) & 16(16\epsilon^2 + 1) & -4(12\epsilon + 1)\\
-64\epsilon & -4(12\epsilon + 1) & 400\epsilon^2 + 24\epsilon + 1
\end{pmatrix}\,.
\end{equation}
Its non-trivial eigenvalues are identical and read
\begin{equation}
Z^2\lambda_{1,2}^{\star} = -\left(3 + \frac{1}{4\epsilon}\right)H\,,
\end{equation}
so that we find the following stability properties. For \(\epsilon > 0\) or \(\epsilon < -\frac{1}{12}\), this point is an attractor for \(H > 0\) and a repeller for \(H < 0\). For \(-\frac{1}{12} < \epsilon < 0\), the opposite is the case. Finally, for \(\epsilon = -\frac{1}{12}\), it becomes a non-hyperbolic projective fixed point. The radial dynamics is given by
\begin{equation}
\dot{\vek{z}} = \frac{H}{4\epsilon}\vek{z}\,,
\end{equation}
so that for \(\epsilon > 0\) one obtains a future finite time singularity for \(H > 0\) and a past finite time singularity for \(H < 0\), while for \(\epsilon < 0\) the opposite behavior is found. Also in this case the barotropic index becomes
\begin{equation}
w_{\lambda} = -1 - \frac{1}{6\epsilon}\,.
\end{equation}

\item
For \(\epsilon < 0\), one finds another pair of projective fixed points given by
\begin{equation}\label{eq:profix234}
K = 0\,, \quad
L = \left(3 \pm \sqrt{-\frac{3}{\epsilon}}\right)H\,.
\end{equation}
The Jacobi matrix now becomes
\begin{equation}
Z^2\mat{J}^{\star} = \frac{\left(6\sqrt{-\epsilon} \pm \sqrt{3}\right)}{\sqrt{-\epsilon}\left(-10\epsilon \pm 6\sqrt{-3\epsilon} + 3\right)}H\begin{pmatrix}
\left(3\sqrt{-\epsilon} \pm \sqrt{3}\right)^2 & 0 & 3\epsilon \mp \sqrt{-3\epsilon}\\
0 & -10\epsilon \pm 6\sqrt{-3\epsilon} + 3 & 0\\
3\epsilon \mp \sqrt{-3\epsilon} & 0 & -\epsilon
\end{pmatrix}
\end{equation}
and possesses two identical eigenvalues
\begin{equation}
Z^2\lambda_{1,2}^{\star} = \left(6 \pm \sqrt{-\frac{3}{\epsilon}}\right)H\,.
\end{equation}
We thus see that the point given by the upper sign is always a repeller for \(H > 0\) and an attractor for \(H < 0\). The same applies to the lower sign for \(\epsilon < -\frac{1}{12}\), but the opposite is the case for \(-\frac{1}{12} < \epsilon < 0\), while at \(\epsilon = -\frac{1}{12}\) one finds a non-hyperbolic projective fixed point. In all cases, the radial dynamics is given by
\begin{equation}
\dot{\vek{z}} = -3H\vek{z}\,,
\end{equation}
so that one has past finite time singularities for \(H > 0\) and future finite time singularities for \(H < 0\). The barotropic index~\eqref{eq:baroin3} takes the stiff fluid value \(w_{\lambda} = 1\).

\item
Finally, for \(\epsilon < 0\) one also finds a pair of projective fixed points at
\begin{equation}\label{eq:profix235}
K = \left(\frac{3}{2} \pm \sqrt{-\frac{3}{2\epsilon}}\right)H\,, \quad
L = -3H\,,
\end{equation}
whose Jacobi matrix takes the lengthy form
\begin{equation}
\begin{split}
Z^2\mat{J}^{\star} &= \frac{\sqrt{3}\left(\sqrt{3} \pm 6\sqrt{-\epsilon}\right)}{\epsilon\left(-196\epsilon \pm 12\sqrt{-3\epsilon} + 3\right)\left(-68\sqrt{3}\epsilon \pm 36\sqrt{-\epsilon} + 3\sqrt{3}\right)}H\\
&\phantom{=}\cdot \left(\begin{array}{c}
9\epsilon\left(-68\epsilon \pm 12\sqrt{-3\epsilon} + 3\right)\left(-36\sqrt{3}\epsilon \pm 4\sqrt{-\epsilon} - 5\sqrt{3}\right)\\
4(8\epsilon + 3)\sqrt{-3\epsilon}\left(6\sqrt{-\epsilon} \pm \sqrt{3}\right)\left(-68\epsilon \pm 12\sqrt{-3\epsilon} + 3\right)\\
\left(-68\epsilon \pm 12\sqrt{-3\epsilon} + 3\right)\left(-204\sqrt{3}\epsilon^2 \mp 108(-\epsilon)^{3/2} - 43\sqrt{3}\epsilon \pm 36\sqrt{-\epsilon} + 3\sqrt{3}\right)
\end{array}\right.,\\
&\phantom{=}\begin{array}{c}
24(-\epsilon)^{3/2}\left(18\sqrt{-3\epsilon} \mp 7\right)\left(-68\epsilon \pm 12\sqrt{-3\epsilon} + 3\right) \\
16(3 + 16\epsilon)\left(-68\sqrt{3}\epsilon \pm 36\sqrt{-\epsilon} + 3\sqrt{3}\right)\\
4\sqrt{-\epsilon}\left(-68\epsilon \pm 12\sqrt{-3\epsilon} + 3\right)\left(68\sqrt{3}(-\epsilon)^{3/2} \mp 2\epsilon - 6\sqrt{-3\epsilon} \mp 3\right)
\end{array},\\
&\phantom{=}\left.\begin{array}{c}
-\epsilon\left(-68\epsilon \pm 12\sqrt{-3\epsilon} + 3\right)\left(-108\sqrt{3}\epsilon \pm 12\sqrt{-\epsilon} + \sqrt{3}\right)\\
32(-\epsilon)^{3/2}\left(2\sqrt{-3\epsilon} \pm 1\right)\left(-68\epsilon \pm 12\sqrt{-3\epsilon} + 3\right)\\
-\epsilon\left(-68\epsilon \pm 12\sqrt{-3\epsilon} + 3\right)\left(-68\sqrt{3}\epsilon \pm 36\sqrt{-\epsilon} + 3\sqrt{3}\right)
\end{array}\right)\,,
\end{split}
\end{equation}
and possesses the eigenvalues
\begin{equation}
Z^2\lambda_1^{\star} = \left(6 \pm \sqrt{-\frac{3}{\epsilon}}\right)H\,, \quad
Z^2\lambda_2^{\star} = -\left(6 \pm \sqrt{-\frac{3}{\epsilon}}\right)H\,.
\end{equation}
Since these eigenvalues always have opposite signs, it follows that both points are always saddle points, independently of \(\epsilon\), while the one corresponding to the lower sign becomes non-hyperbolic at \(\epsilon = -\frac{1}{12}\). Finally, the radial dynamics is given by
\begin{equation}
\dot{\vek{z}} = -3H\vek{z}
\end{equation}
so that also these points are past finite time singularities for \(H > 0\) and future finite time singularities for \(H < 0\). Also in this case the barotropic index~\eqref{eq:baroin3} takes the stiff fluid value \(w_{\lambda} = 1\).
\end{enumerate}

\subsection{Phase diagrams}\label{ssec:phase2}
The cosmological dynamics we found for the different branches depending on the value of the parameter \(\epsilon\) can most easily be visualized with the help of phase diagrams, which show the (projective) fixed points, radial dynamics and possible trajectories. Since the dimension of the dynamical system is different for the different branches, we use different methods for drawing these diagrams:

For the first branch, the system is two-dimensional. We can therefore display the whole phase space, which we choose to compactify as a Poincaré disk model, and write the dynamical variables \(H, K\) in polar coordinates \(r, \phi\) as
\begin{equation}
H = \frac{r\cos\phi}{\sqrt{1 - r^2}}\,, \quad
K = \frac{r\sin\phi}{\sqrt{1 - r^2}}\,,
\end{equation}
where \(r < 1\). In this model, projective fixed points appear as radial lines through the origin, and they form separatrices of the phase space. We display these likes as solid for attractors, dashed for repellers and dotted for non-hyperbolic projective fixed points. Finally, we draw them in blue for \(\dot{Z} < 0\), black for \(\dot{Z} = 0\) and red for \(\dot{Z} > 0\).

For the second and third branch, we have a three-dimensional system. It is helpful to recall that we can separate the angular and radial dynamics as shown in section~\ref{ssec:prodecg}. Since the angular dynamics given by \(\dot{\vek{n}}\) as a function of \(\vek{n}\) can be considered independently from the radial dynamics, we can display it using a two-dimensional representation of the unit sphere, which contains the points \(\vek{n}\).
Finally, due to the reflection symmetry discussed in section~\ref{ssec:prodecg}, it suffices to display one hemisphere, which we choose as \(H \geq 0\). We then use spherical coordinates \((\theta, \phi)\) such that
\begin{equation}
H = \cos\theta\,, \quad
K = \sin\theta\cos\phi\,, \quad
L = \sin\theta\sin\phi\,,
\end{equation}
and display \(\theta\) on the radial axis and \(\phi\) on the polar axis. In these plots, projective fixed points appear as single points in the projected phase space, and we introduce the following symbols to mark these points and denote their stability properties, which follows from the signs of the real parts of the non-trivial eigenvalues (corresponding to angular directions, i.e., excluding the vanishes radial eigenvalue) of the Jacobi matrix:
\begin{enumerate}
\item
\(\otimes\): all real parts vanish.
\item
\(\circ\): there are vanishing and positive real parts.
\item
\(\bullet\): there are vanishing and negative real parts.
\item
\(\boxtimes\): there are positive and negative real parts.
\item
\(\square\): there are only positive real parts.
\item
\(\blacksquare\): there are only negative real parts.
\end{enumerate}
These symbols are shown in blue for \(\dot{Z} < 0\), black for \(\dot{Z} = 0\) and red for \(\dot{Z} > 0\). Due to the complicated structure of the phase space in these models, we use colors to distinguish different regions or the phase space, which are separated by separatrices, so that all trajectories within each colored region have the same qualitative dynamics. A detailed analysis of the phase space for all possible parameter values is given below.

\subsubsection{Branch 1}\label{sssec:phase21}
We start with the first branch. Based on the previous analysis of the existence and stability of projective fixed points, we find that there are four different cases, which exhibit qualitatively different behavior.
\begin{figure}[tp]
\includegraphics[width=\textwidth]{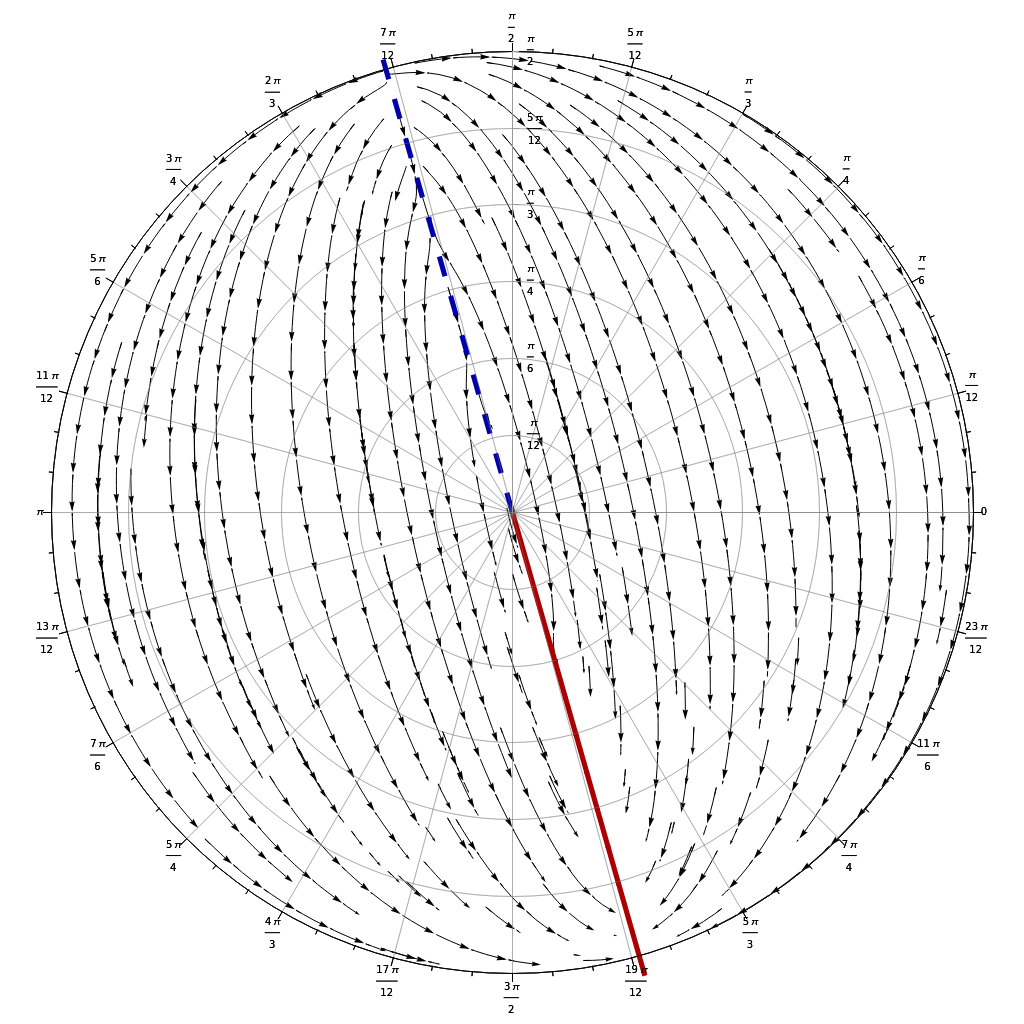}
\caption{Phase diagram of the first branch in theories of type 2 with \(\epsilon = 1\).}
\label{fig:phase21p1}
\end{figure}
\begin{figure}[tp]
\includegraphics[width=\textwidth]{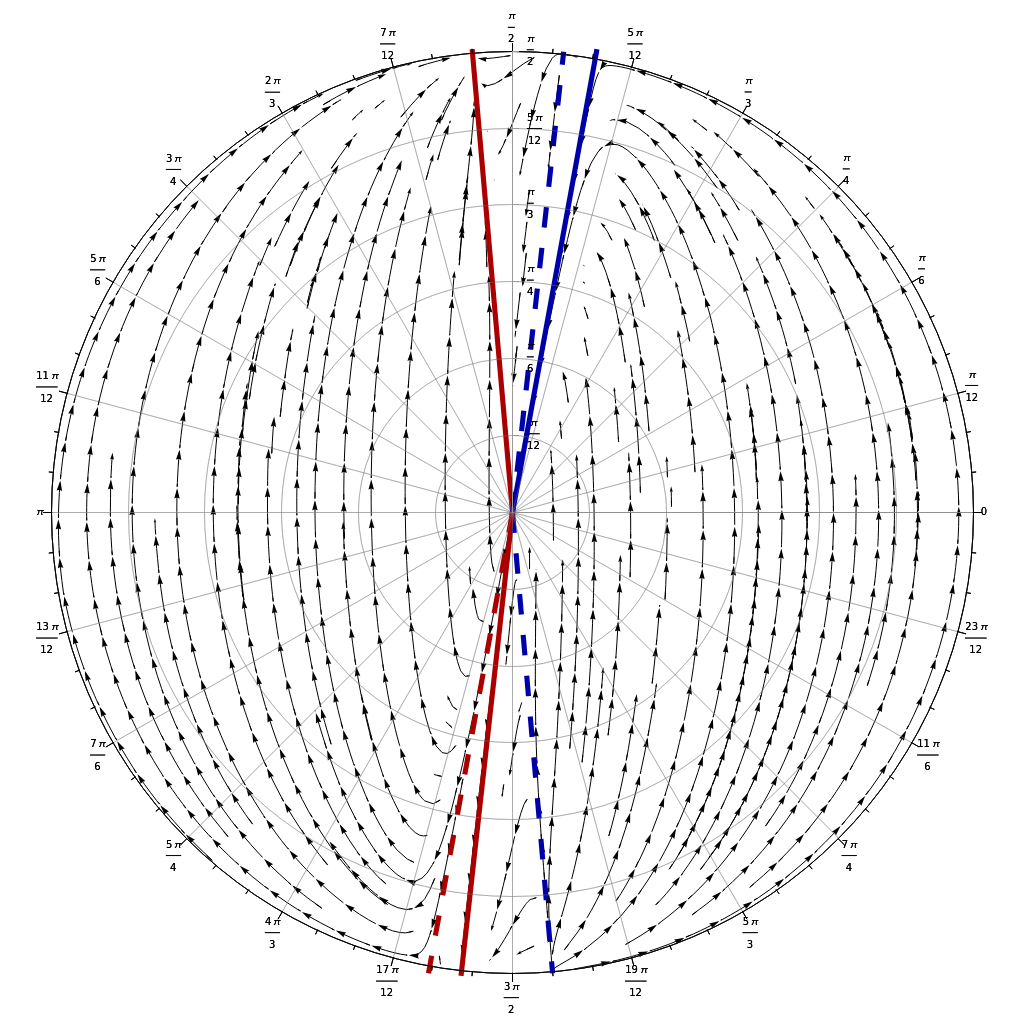}
\caption{Phase diagram of the first branch in theories of type 2 with \(\epsilon = -\frac{1}{24}\).}
\label{fig:phase21m1_24}
\end{figure}
\begin{figure}[tp]
\includegraphics[width=\textwidth]{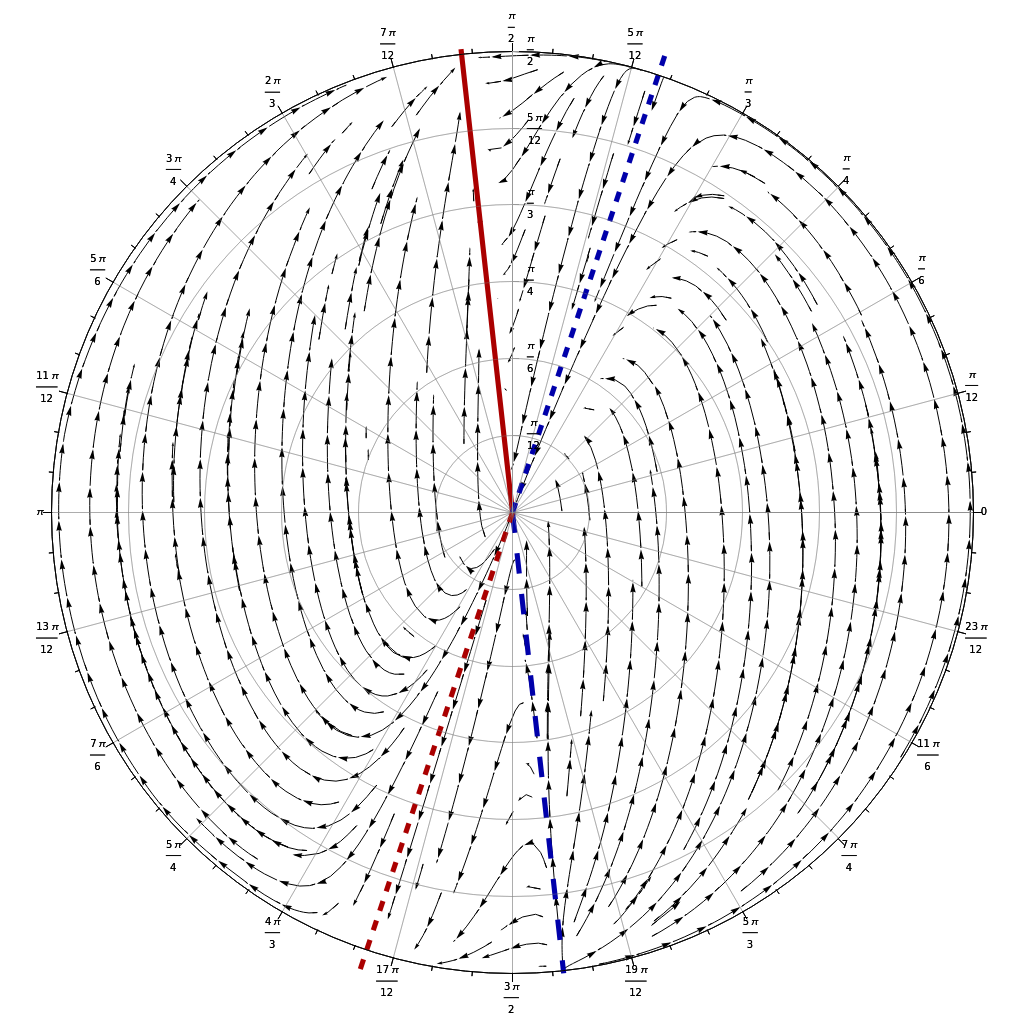}
\caption{Phase diagram of the first branch in theories of type 2 with \(\epsilon = -\frac{1}{12}\).}
\label{fig:phase21m1_12}
\end{figure}
\begin{figure}[tp]
\includegraphics[width=\textwidth]{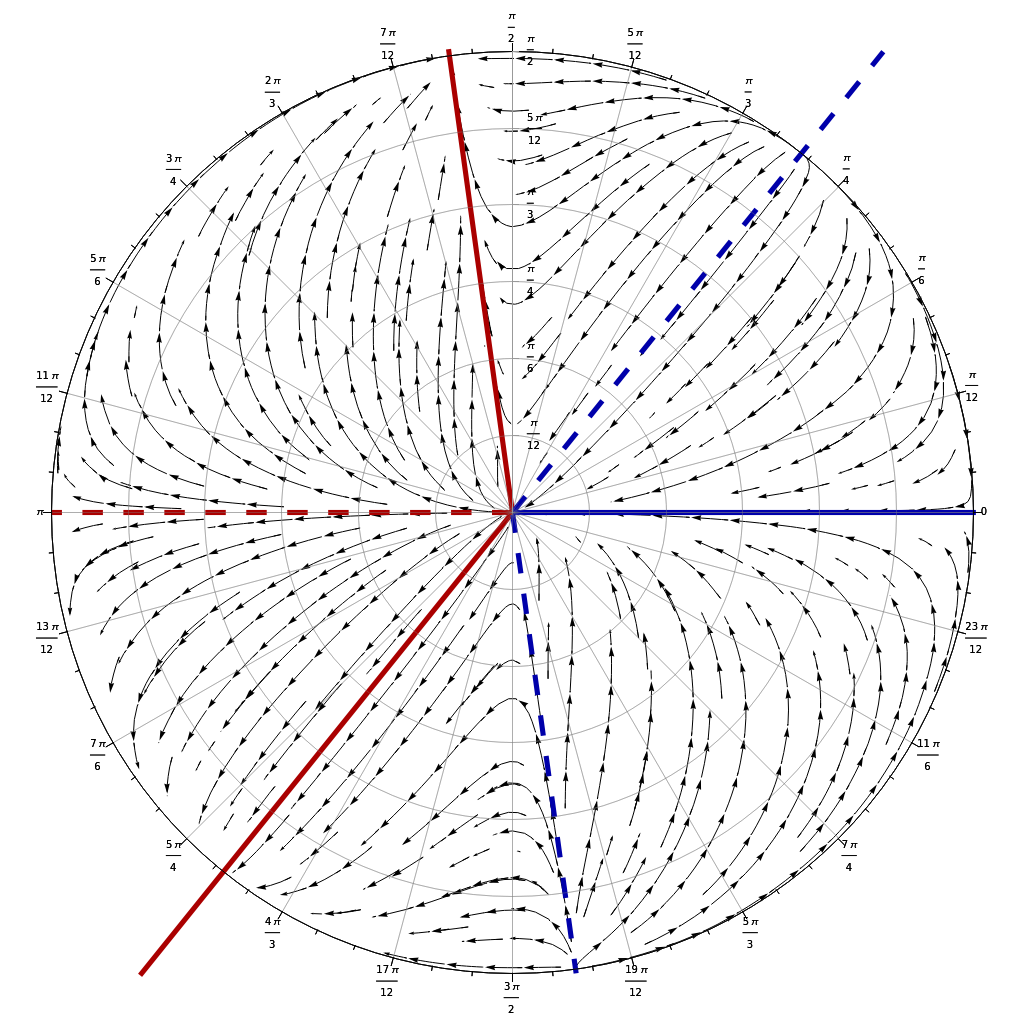}
\caption{Phase diagram of the first branch in theories of type 2 with \(\epsilon = -\frac{1}{6}\).}
\label{fig:phase21m1_6}
\end{figure}
\begin{enumerate}
\item
First, we consider the case \(\epsilon > 0\), which is depicted in figure~\ref{fig:phase21p1} for \(\epsilon = 1\). In this case there is only one projective fixed point~\eqref{eq:profix211}, as discussed in section~\ref{sssec:profixp21}. This is an attractor for \(H > 0\) and a repeller for \(H < 0\). Almost all trajectories originate from a past finite time singularity, undergo initial contraction, followed by a bounce and an expanding phase, which leads to a future finite time singularity. The only exceptions are the trajectories corresponding to the projective fixed point itself, which either approach or originate from the origin \(H = K = 0\), but do not cross it, as it is a fixed point.

\item
We then consider the case \(-\frac{1}{12} < \epsilon < 0\) shown in figure~\ref{fig:phase21m1_24}, where we chose \(\epsilon = -\frac{1}{24}\). In this case also the projective fixed point pair~\eqref{eq:profix212} appears. For \(H > 0\), corresponding to the right half of the figure, one of these projective fixed points is an attractor with \(\dot{Z} < 0\), while the other two are repellers, which correspond to past finite time (big bang) singularities. For \(H < 0\), the opposite is the case, and we find one repeller and two attractive future finite time (big crunch) singularities.

We now find different types of trajectories. There are trajectories which are confined to the right half \(H > 0\), which originate from one of the repulsive past finite time (big bang) singularities and then approach the attractive projective fixed point, where they develop into an eternally expanding universe, whose Hubble parameter asymptotically approaches \(H \to 0\). The time-reversed case can be found in the left half of the figure, where trajectories emanate from an eternally contracting universe and approach a future finite time (big crunch) singularity. Finally, there are also trajectories, which originate from a big bang singularity, undergo a turnaround at \(H = 0\) and lead to a big crunch singularity.
\item
A qualitative change of behavior occurs at \(\epsilon = -\frac{1}{12}\), which is shown in figure~\ref{fig:phase21m1_12}. Here two projective fixed points coincide, and merge into a singular non-hyperbolic projective fixed point. For each sign of \(H\), this combines what were previously one repeller and one attractor, and so it attracts trajectories on one side, while repelling them on the other side. There still remains one hyperbolic projective fixed point, which is repulsive with \(\dot{Z} < 0\) for \(H > 0\) and attractive with \(\dot{Z} > 0\) for \(H < 0\).

The possible trajectories in this phase space are similar to the previous case. Trajectories emanating from an expanding past finite time (big bang) singularity can either remain eternally expanding or undergo a turnaround and approach a future finite time (big crunch) singularity. The latter is also approached by trajectories which were eternally contracting in the past.

\item
Finally, the case \(\epsilon < -\frac{1}{12}\) is shown in figure~\ref{fig:phase21m1_6} with \(\epsilon = -\frac{1}{6}\). In this case, the behavior of the system is qualitatively the same as in the case \(-\frac{1}{12} < \epsilon < 0\), with the only difference that the roles of two projective fixed points are reversed.
\end{enumerate}

\subsubsection{Branch 2}\label{sssec:phase22}
We then come to the phase diagrams for the second branch. In this case we must distinguish eight different cases, each of which exhibits a different qualitative behavior.
\begin{figure}[tp]
\includegraphics[width=\textwidth]{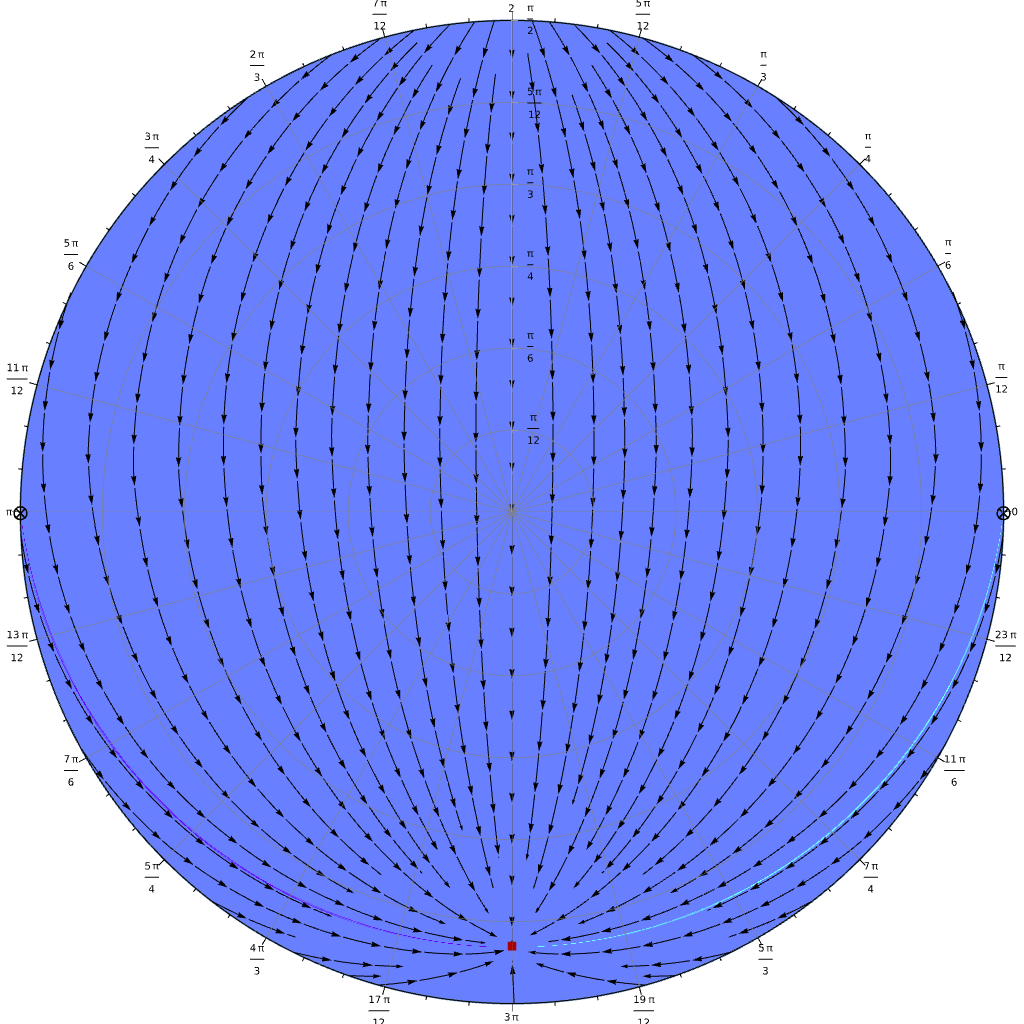}
\caption{Phase diagram of the second branch in theories of type 2 with \(\epsilon = 1\).}
\label{fig:phase22p1}
\end{figure}
\begin{figure}[tp]
\includegraphics[width=\textwidth]{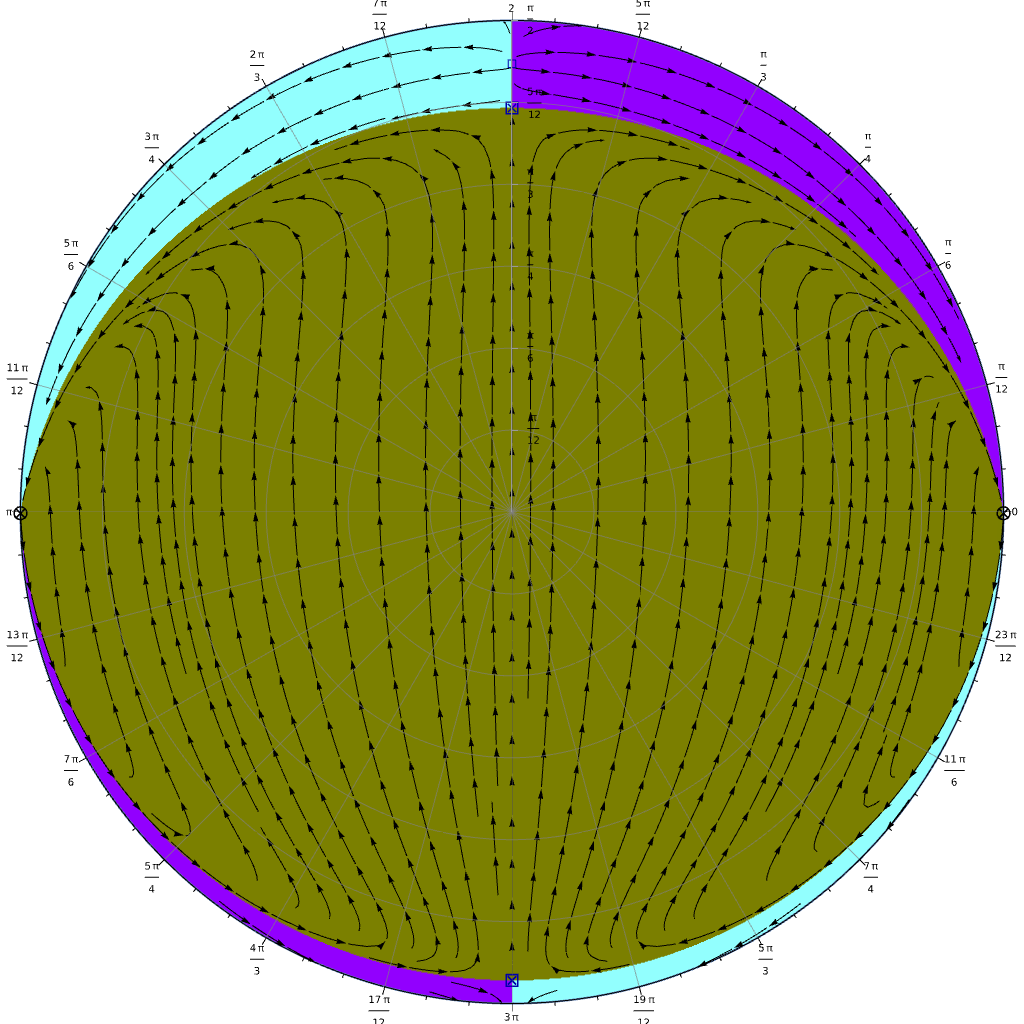}
\caption{Phase diagram of the second branch in theories of type 2 with \(\epsilon = -\frac{1}{24}\).}
\label{fig:phase22m1_24}
\end{figure}
\begin{figure}[tp]
\includegraphics[width=\textwidth]{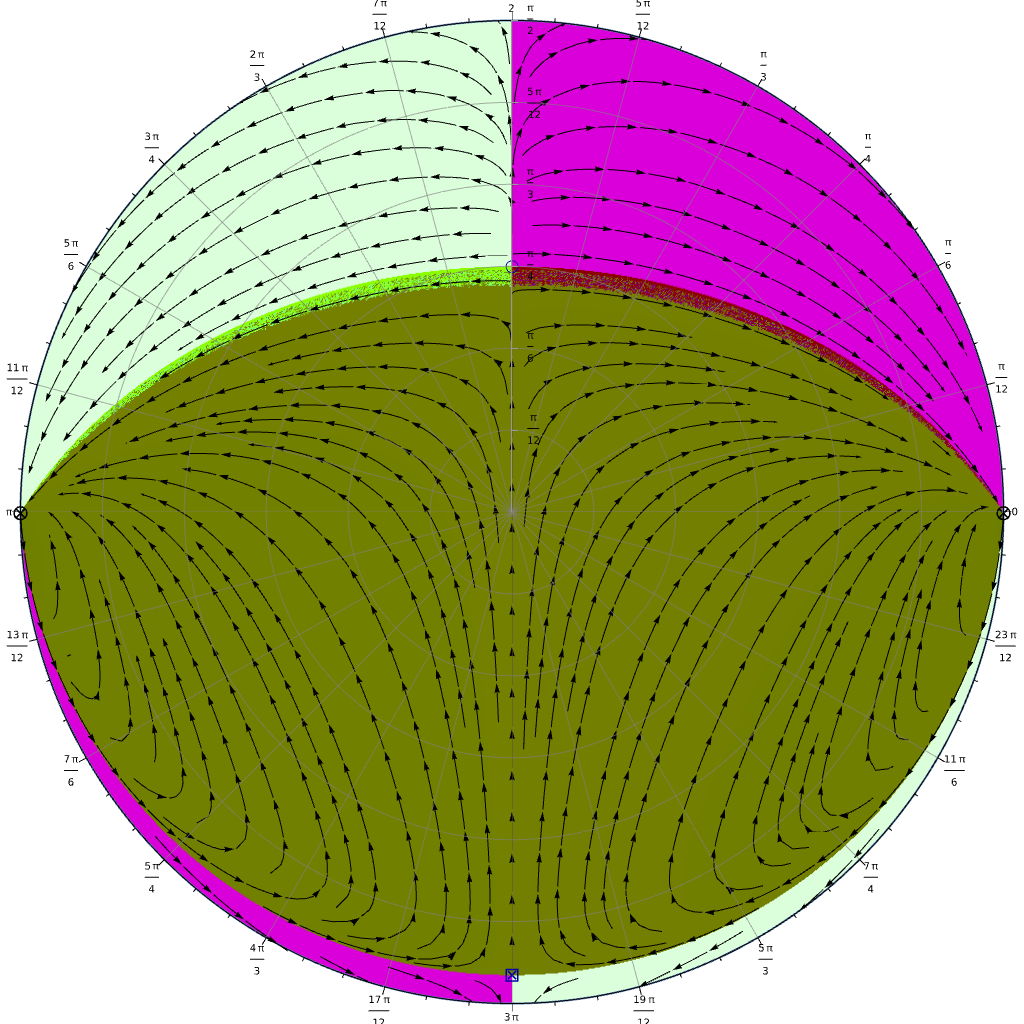}
\caption{Phase diagram of the second branch in theories of type 2 with \(\epsilon = -\frac{1}{12}\).}
\label{fig:phase22m1_12}
\end{figure}
\begin{figure}[tp]
\includegraphics[width=\textwidth]{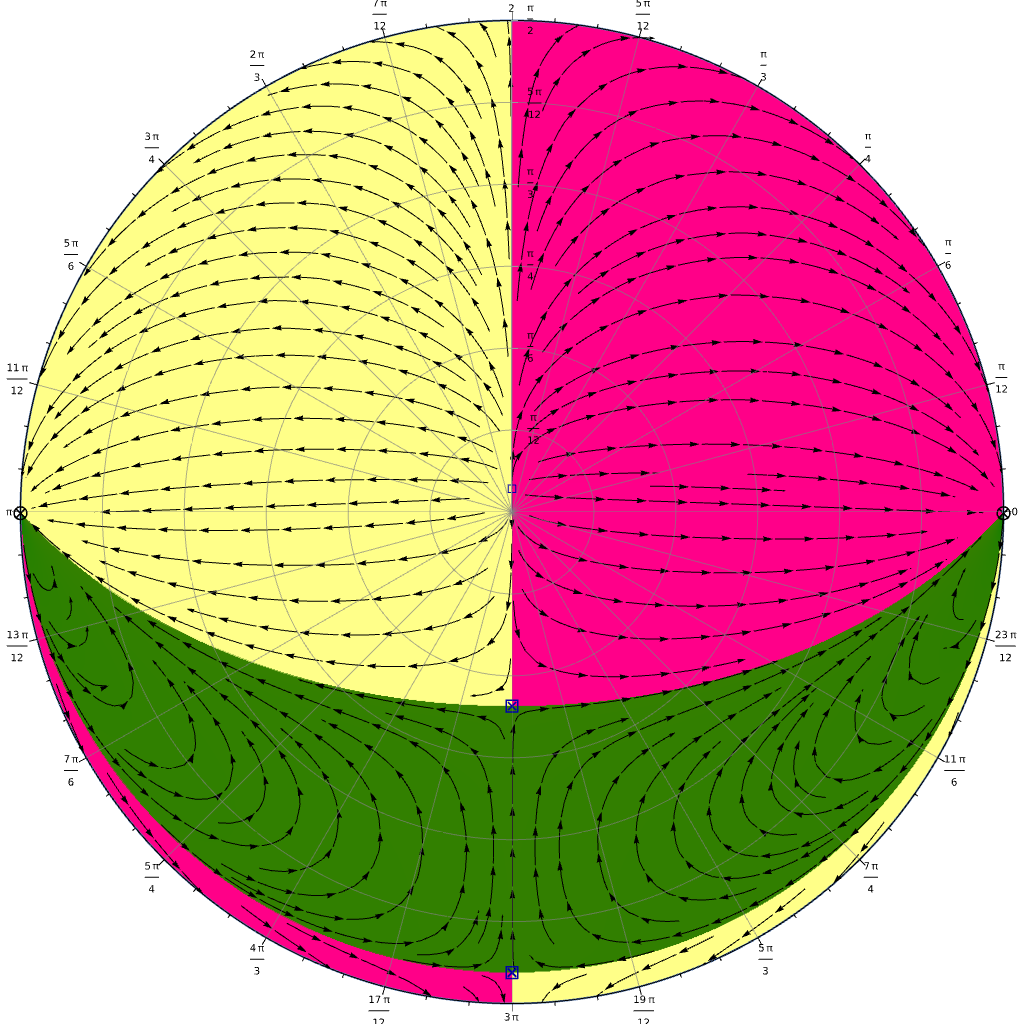}
\caption{Phase diagram of the second branch in theories of type 2 with \(\epsilon = -\frac{7}{60}\).}
\label{fig:phase22m7_60}
\end{figure}
\begin{figure}[tp]
\includegraphics[width=\textwidth]{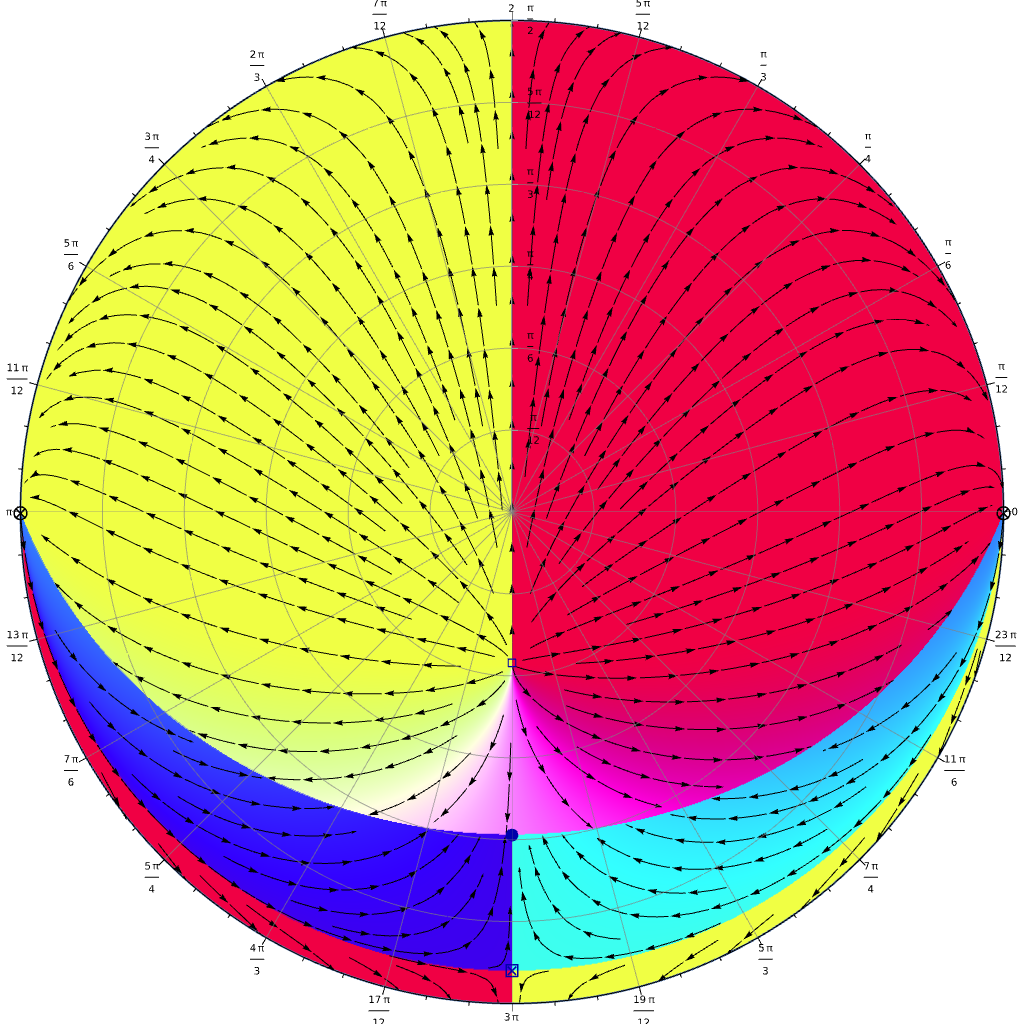}
\caption{Phase diagram of the second branch in theories of type 2 with \(\epsilon = -\frac{3}{20}\).}
\label{fig:phase22m3_20}
\end{figure}
\begin{figure}[tp]
\includegraphics[width=\textwidth]{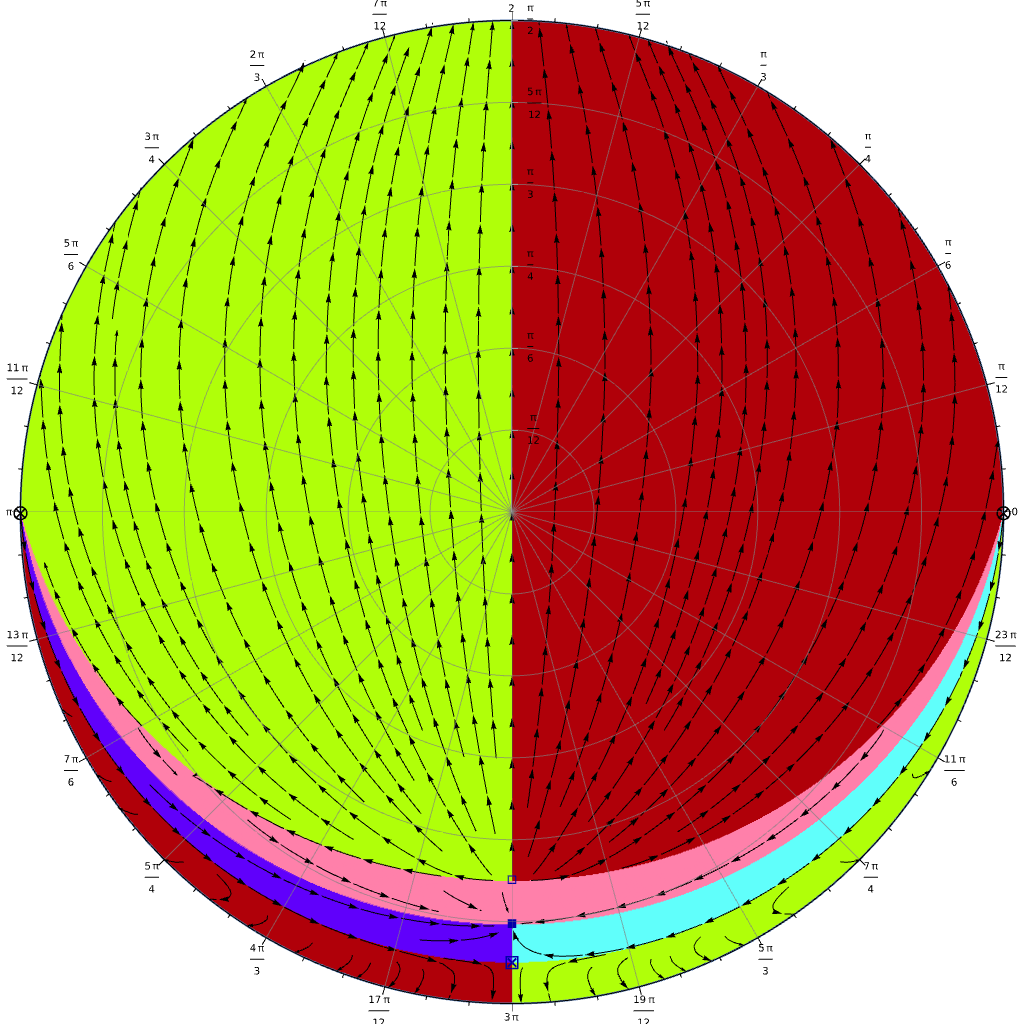}
\caption{Phase diagram of the second branch in theories of type 2 with \(\epsilon = -\frac{9}{20}\).}
\label{fig:phase22m9_20}
\end{figure}
\begin{figure}[tp]
\includegraphics[width=\textwidth]{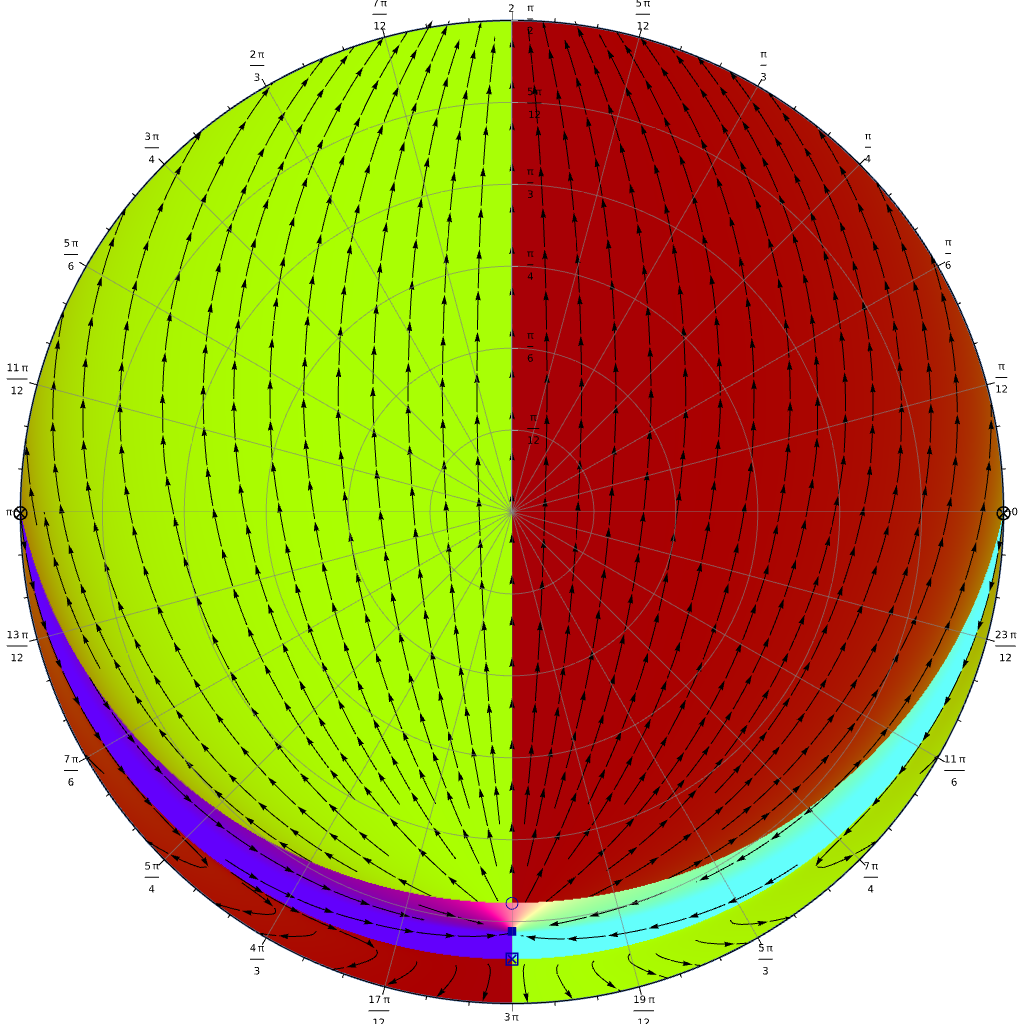}
\caption{Phase diagram of the second branch in theories of type 2 with \(\epsilon = -\frac{3}{4}\).}
\label{fig:phase22m3_4}
\end{figure}
\begin{figure}[tp]
\includegraphics[width=\textwidth]{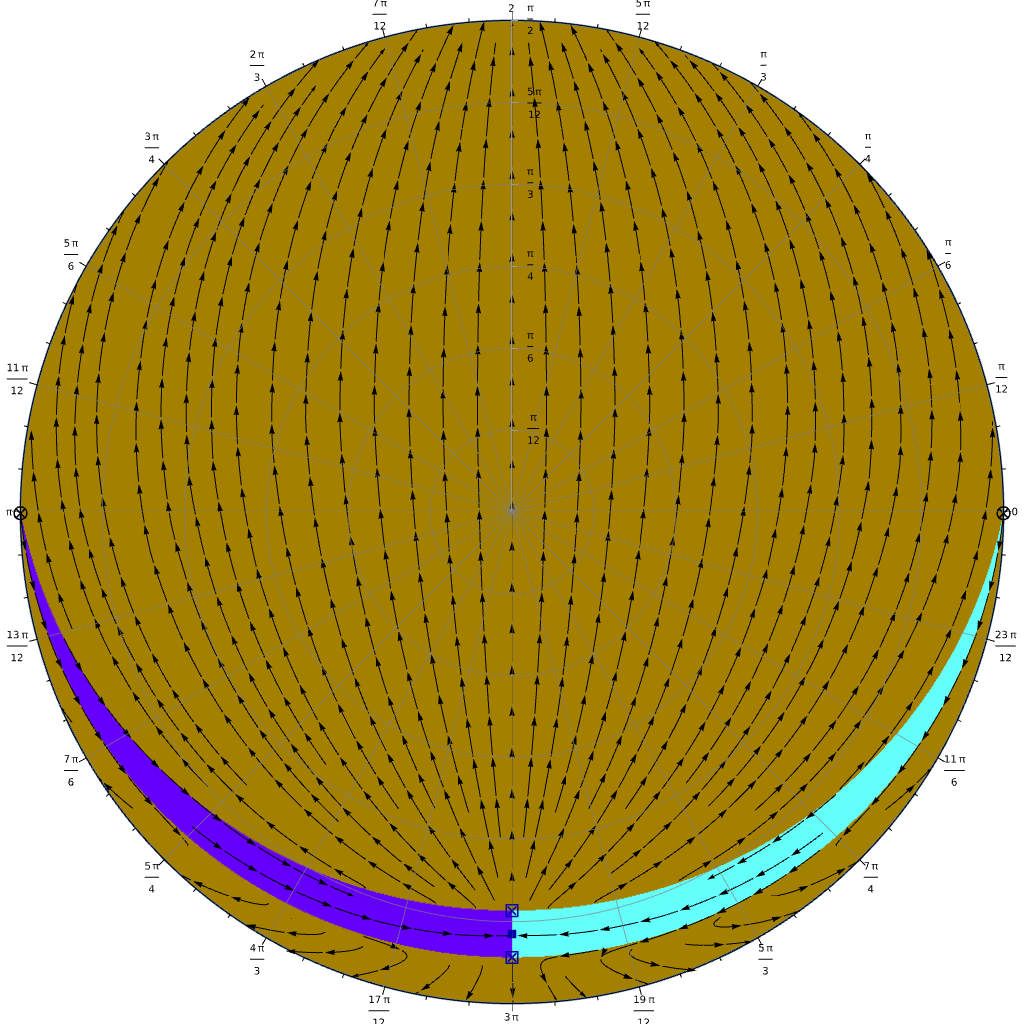}
\caption{Phase diagram of the second branch in theories of type 2 with \(\epsilon = -1\).}
\label{fig:phase22m1}
\end{figure}
\begin{enumerate}
\item
We start with the case \(\epsilon > 0\), which is displayed in figure~\ref{fig:phase22p1}, where we have chosen \(\epsilon = 1\). Recall from section~\ref{sssec:profixp22} that the projective fixed point~\eqref{eq:profix221} always exists, independently of \(\epsilon\), and that it is always a non-hyperbolic fixed point with vanishing radial dynamics, \(\dot{Z} = 0\). In the figure, it appears twice on the perimeter, which represents the equatorial plane \(H = 0\), where it intersects the central horizontal line \(L = 0\) at \(K = \pm 1\) on the right and left, indicated by the symbols \(\otimes\). Also the projective fixed point~\eqref{eq:profix222} always exists and appears in the diagram, which shows only the upper hemisphere \(H \geq 0\), on the lower half \(L < 0\) of the central vertical line \(K = 0\), marked by the symbol \(\blacksquare\). It is an attractor with \(\dot{Z} > 0\), and thus an attractive future finite time singularity.

Further, recall from section~\ref{ssec:bt2} that only bounces are possible for \(\epsilon > 0\), which can be seen by the fact that all trajectories pass the perimeter in the inward direction, i.e., from \(H < 0\) towards \(H > 0\). They originate from the projective fixed point~\eqref{eq:profix222} on the other side \(H < 0\) which is not displayed here, and which is a repulsive past finite time singularity. Hence, they represent a universe which is initially contracting after a finite time singularity, then bounces, expands and ends in another finite time singularity. The only exceptional trajectories are the stable and unstable manifolds of the projective fixed points~\eqref{eq:profix221}, which connect these points to the attractor and repeller of the system.

\item
We then consider the case \(-\frac{1}{12} < \epsilon < 0\), which is represented by setting \(\epsilon = -\frac{1}{24}\) in figure~\ref{fig:phase22m1_24}. The projective fixed point~\eqref{eq:profix221} is unchanged compared to the previous case, and will remain unchanged for all further parameter values we consider here. The point~\eqref{eq:profix222}, which is now shown on the top of the central vertical line \(K = 0\), has become a repulsive past finite time singularity, indicated by the symbol \(\square\). Further, we now find the pair~\eqref{eq:profix223} of projective fixed points, both of which are saddle points, and which are likewise located along the central vertical line \(K = 0\) marked by \(\boxtimes\), and correspond to past finite time singularities. Only turnarounds are possible for \(\epsilon < 0\), which is seen by the fact that now all trajectories point outward at the perimeter \(H = 0\).

The saddle points of the system are connected by separatrices, which divide the phase space into different region, which contain qualitatively different trajectories. Trajectories emanating from the past finite time singularity~\eqref{eq:profix222} with \(H > 0\) represent an initially expanding universe, which may undergo a turnaround if they cross the equatorial plane \(H = 0\), before approaching the point~\eqref{eq:profix221}. On the opposite hemisphere, one finds corresponding trajectories emanating from these saddle points, which approach the future finite time singularity~\eqref{eq:profix222} with \(H < 0\) not displayed in the diagram. Another class of trajectories is confined to the area between the displayed saddle points. These trajectories emanate from one of the saddle points~\eqref{eq:profix221}, keeping either \(K > 0\) or \(K < 0\), while \(L\) changes its sign once along the trajectory from \(L < 0\) to \(L > 0\), and then approach the same saddle point again. Along these trajectories, the universe is eternally expanding in past and future, since \(H\) remains positive. An analogous type of trajectories for an eternally contracting universe exists on the other hemisphere \(H < 0\). Further, the different saddle points~\eqref{eq:profix223} are connected by an unstable trajectory, which starts from an expanding past finite time singularity and proceeds towards an eternally slowing expansion. An analogous trajectory exists for the contracting universe \(H < 0\), from a past eternal slow contraction towards a future finite time singularity.

\item
The case \(\epsilon = -\frac{1}{12}\) is shown in figure~\ref{fig:phase22m1_12}. Now the projective fixed points represented by the solution~\eqref{eq:profix222} and the upper sign of the solution~\eqref{eq:profix223} coincide, and this common point becomes non-hyperbolic, as shown by the symbol \(\circ\) in the upper half \(L > 0\) of the central vertical line \(K = 0\). The lower sign of the solution~\eqref{eq:profix223} remains a saddle point, which appears on the lower half \(L < 0\) of the line \(K = 0\), and will remain there also for the remaining parameter values. Both correspond to past finite time singularities. We find the same types of trajectories as in the previous case, but the merged projective fixed point is now located at the intersection of two separatrices.

\item
Next, we come to the case \(-\frac{3}{20} < \epsilon < -\frac{1}{12}\), which is represented by choosing the value \(\epsilon = -\frac{7}{60}\) in figure~\ref{fig:phase22m7_60}. Now the projective fixed point~\eqref{eq:profix222} has turned into a saddle point, which is located in the middle of the lower half \(L < 0\) of the central vertical line \(K = 0\). At the same time, the point given by the upper sign in the solution~\eqref{eq:profix223} is now a repeller, which is found close to the center of the diagram. The qualitative behavior is similar to the previously discussed case \(-\frac{1}{12} < \epsilon < 0\), but two of the projective fixed points have interchanged their roles and positions in the phase space.

\item
Another remarkable case is given for the parameter value \(\epsilon = -\frac{3}{20}\), shown in figure~\ref{fig:phase22m3_20}. We now find that the fixed point~\eqref{eq:profix222} becomes non-hyperbolic, now displayed by the symbol \(\bullet\) in the lower half \(L < 0\) of the central vertical line \(K = 0\). It turns out that the direction of the flow on the separatrix connecting this projective fixed point to the point~\eqref{eq:profix221} changes its sign above and below this critical value of \(\epsilon\), and that the flow stops at this value itself, i.e., all points along the separatrix, which is now located at \(5H + 3L = 0\), become non-hyperbolic fixed points, which are attractive for \(H > 0\) and repulsive for \(H < 0\).

In this case we find that trajectories emanating from the past finite time singularity approach either the non-hyperbolic projective fixed point~\eqref{eq:profix221} or the separatrix constituted by projective fixed points. Trajectories emanating from the fixed point~\eqref{eq:profix221} can also approach this separatrix, or the future finite time singularity for \(H < 0\) on the other hemisphere not shown in the diagram. On the latter one also finds a separatrix constituted by repellers, from which trajectories approach either the future finite time singularity or the non-hyperbolic fixed points.

\item
We then study the case \(-\frac{3}{4} < \epsilon < -\frac{3}{20}\) shown in figure~\ref{fig:phase22m9_20}, where we have chosen \(\epsilon = -\frac{9}{20}\). Now the projective fixed point~\eqref{eq:profix222} is an attractor for \(H > 0\), which again changes the qualitative behavior of trajectories. Trajectories emanating from the past finite time singularity now either approach this new attractor, which is also a past finite time singularity and thus leads to \(Z \to 0\), or the non-hyperbolic fixed points as in the previously discussed cases. The opposite behavior is observed for the future finite time singularities present in the other hemisphere \(H < 0\), which emanate from the now repulsive solution~\eqref{eq:profix222} and either approach a future finite time singularity or the non-hyperbolic fixed points. Trajectories emanating from these non-hyperbolic fixed points approach either attractor for \(H > 0\) or \(H < 0\), and these cases are separated by separatrices connecting the non-hyperbolic and hyperbolic saddle points.

\item
Another important qualitative change occurs at \(\epsilon = -\frac{3}{4}\) shown in figure~\ref{fig:phase22m3_4}. Now the projective fixed point given by the upper sign of the solution~\eqref{eq:profix222} becomes non-hyperbolic. This point is now connected to the non-hyperbolic points~\eqref{eq:profix221} by a separatrix \(3H + L = 0\), which consists entirely of projective fixed points. These points are repulsive for \(H > 0\) and attractive for \(H < 0\). For \(H > 0\) as shown in the diagram, trajectories emanating from the upper side of this separatrix undergo a bounce and approach the non-hyperbolic fixed points~\eqref{eq:profix221}, while trajectories emanating below approach the attractor~\eqref{eq:profix222}. Trajectories emanating from the non-hyperbolic fixed points approach either the attractor~\eqref{eq:profix222} at \(H > 0\) or the attractive separatrix of projective fixed points at \(H < 0\), and these are separated by a separatrix connecting the non-hyperbolic fixed points to the hyperbolic saddle point. Finally, the opposite hemisphere \(H < 0\) contains a repeller~\eqref{eq:profix222} from which trajectories approach the non-hyperbolic fixed points or the attractive separatrix.

\item
Finally, we come to the case \(\epsilon < -\frac{3}{4}\), represented by \(\epsilon = -1\) in figure~\ref{fig:phase22m1}. This case is essentially the opposite of the case \(-\frac{1}{12} < \epsilon < 0\) we discussed before. We find that the point~\eqref{eq:profix222} is an attractive past finite time singularity for \(H > 0\) and a repulsive future finite time singularity for \(H < 0\). The solutions~\eqref{eq:profix223} are saddle points. All saddle points and non-hyperbolic fixed points~\eqref{eq:profix221} are connected by separatrices. Trajectories connect these non-hyperbolic fixed points and the attractor / repeller.
\end{enumerate}

\subsubsection{Branch 3}\label{sssec:phase23}
Finally, we consider the third branch. Here we have four types of qualitatively different phase diagrams which we must distinguish.
\begin{figure}[tp]
\includegraphics[width=\textwidth]{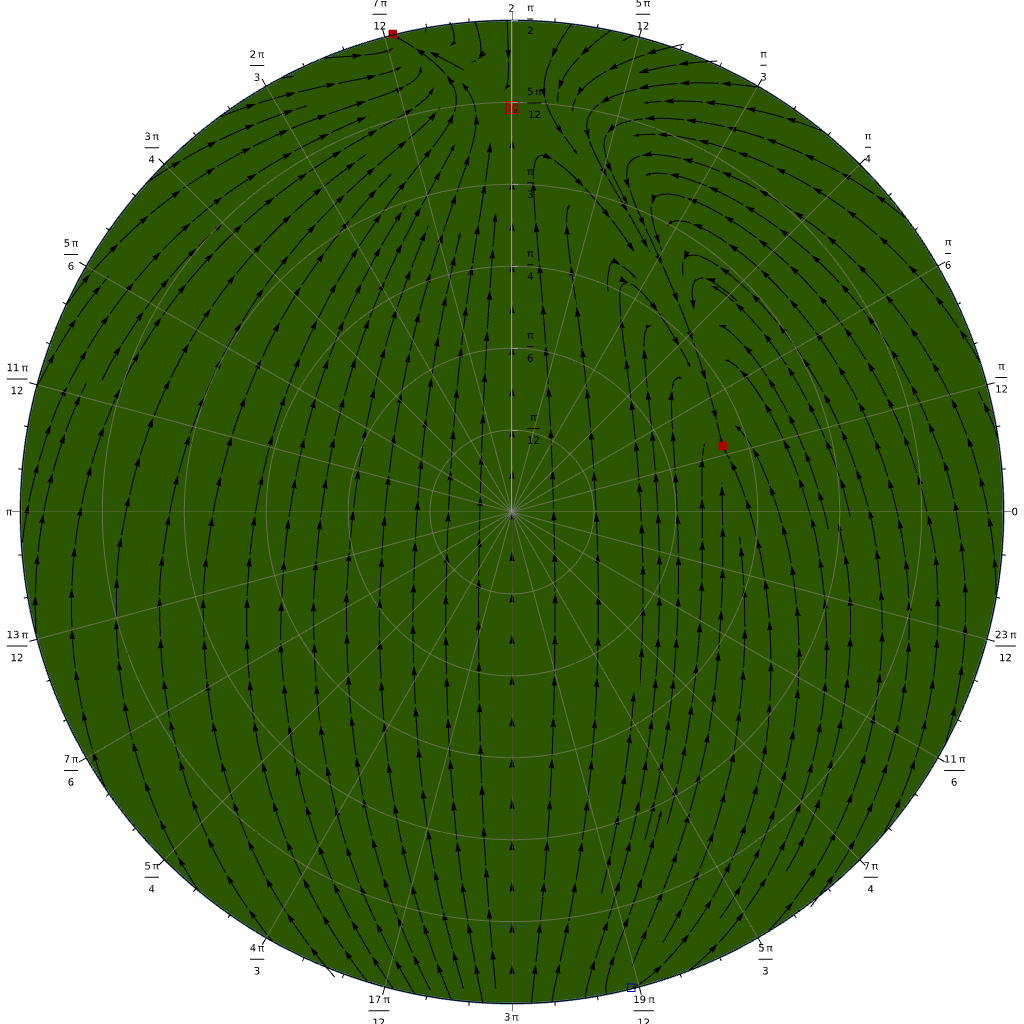}
\caption{Phase diagram of the third branch in theories of type 2 with \(\epsilon = 1\).}
\label{fig:phase23p1}
\end{figure}
\begin{figure}[tp]
\includegraphics[width=\textwidth]{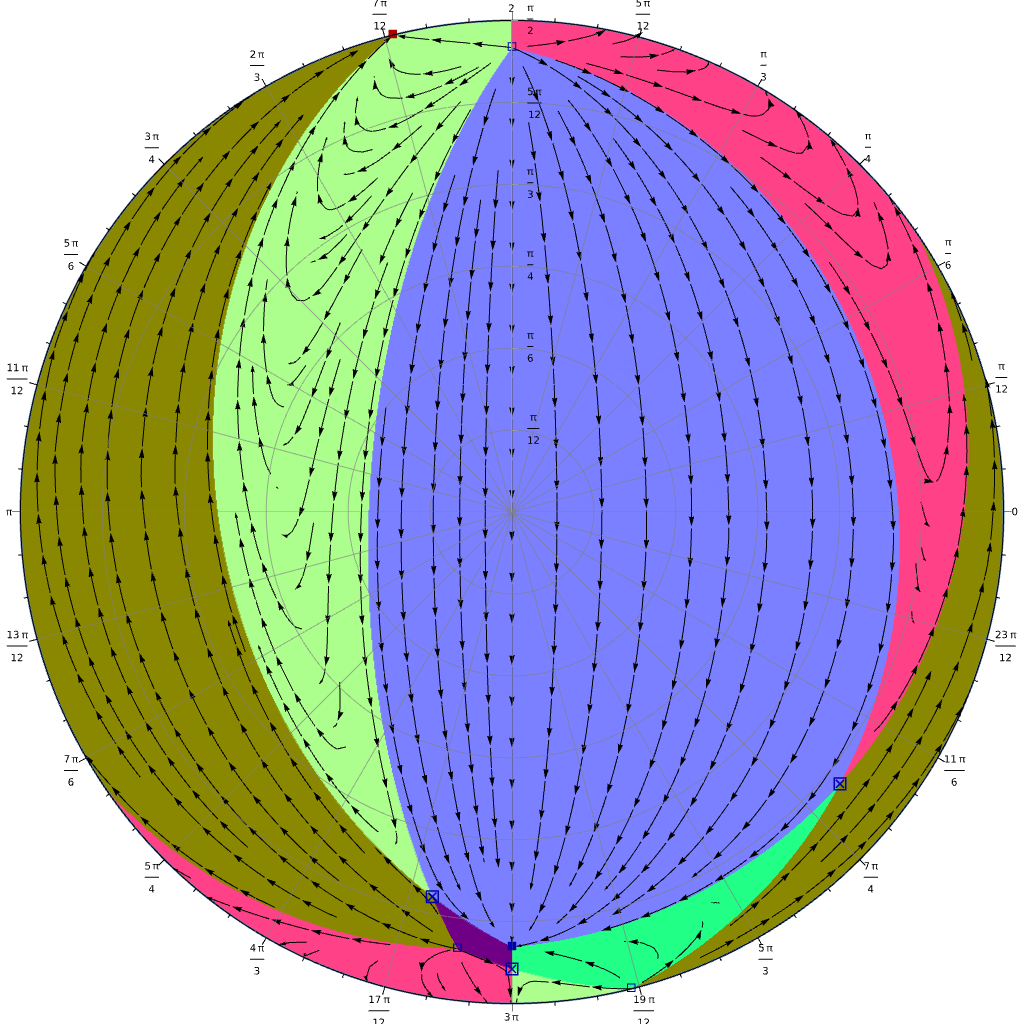}
\caption{Phase diagram of the third branch in theories of type 2 with \(\epsilon = -\frac{1}{24}\).}
\label{fig:phase23m1_24}
\end{figure}
\begin{figure}[tp]
\includegraphics[width=\textwidth]{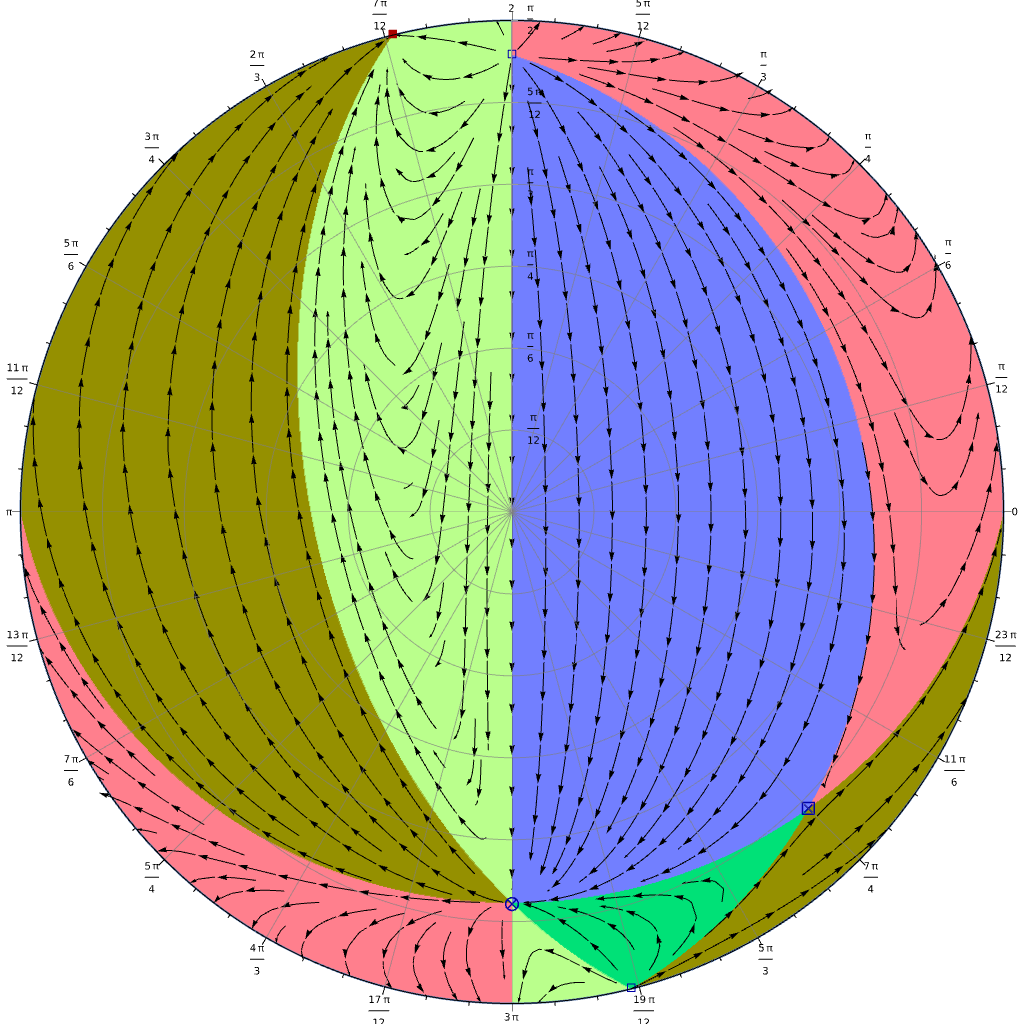}
\caption{Phase diagram of the third branch in theories of type 2 with \(\epsilon = -\frac{1}{12}\).}
\label{fig:phase23m1_12}
\end{figure}
\begin{figure}[tp]
\includegraphics[width=\textwidth]{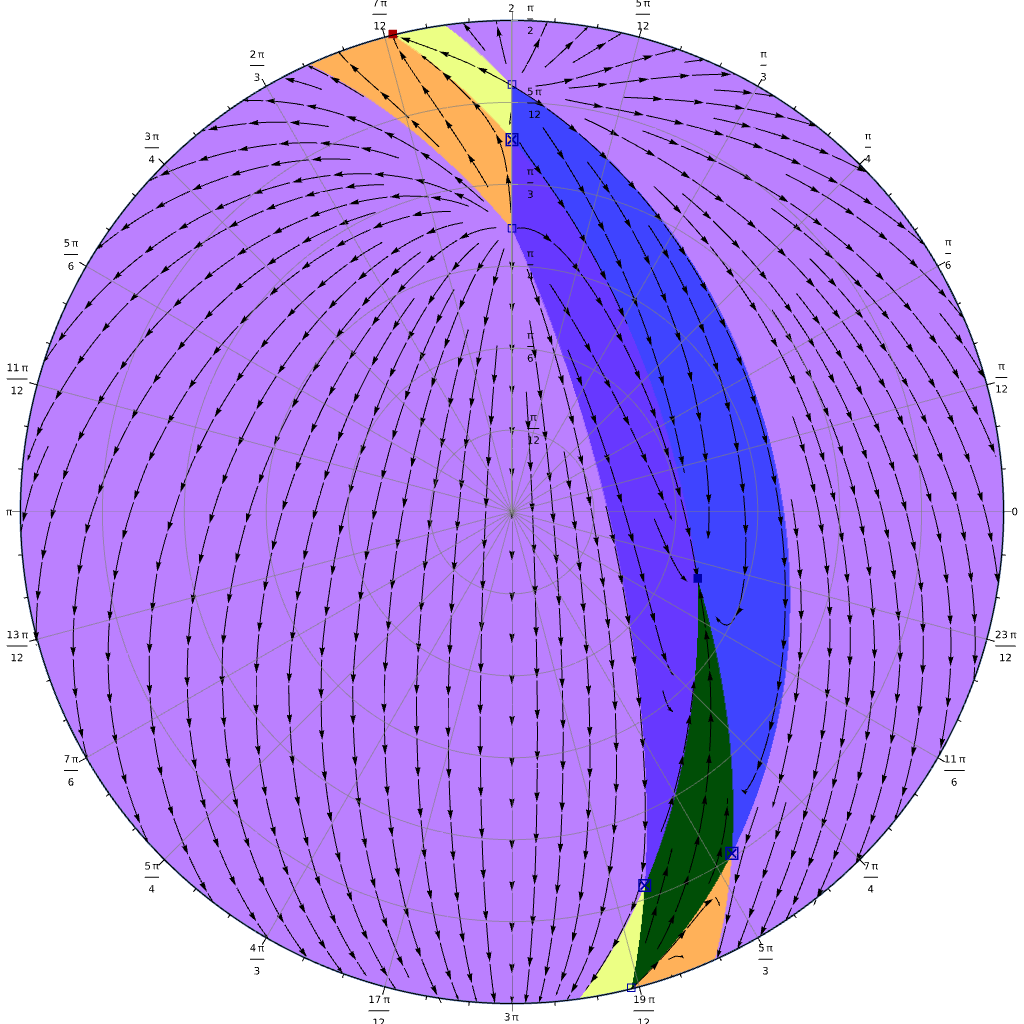}
\caption{Phase diagram of the third branch in theories of type 2 with \(\epsilon = -1\).}
\label{fig:phase23m1}
\end{figure}
\begin{enumerate}
\item
We first consider the case \(\epsilon > 0\), which is displayed in figure~\ref{fig:phase23p1}, where we chose \(\epsilon = 1\). We first take a look at the projective fixed points appearing in this case. The point~\eqref{eq:profix231}, which always exists independently of the value of the parameter \(\epsilon\), is displayed twice at the perimeter of the diagram, which corresponds to the equatorial plane \(H = 0\). It is an attractive future finite time singularity for \(K < 0\), shown close to the upper edge of the diagram, and a repulsive past finite time singularity for \(K > 0\), appearing near the lower edge of the diagram. The projective fixed point~\eqref{eq:profix232}, which is always a saddle point, appears on the central vertical line \(K = 0\) on the upper half \(L > 0\) of the diagram. Finally, we find the projective fixed point~\eqref{eq:profix233} on the right half \(K > 0\) of the diagram. This point is an attractor for \(H > 0\), corresponding to a future finite time singularity. It is remarkable that all projective fixed points are located in the plane
\begin{equation}
4K + L - \left(3 + \frac{1}{2\epsilon}\right) = 0\,.
\end{equation}
Along this plane, the saddle points are connected to the other fixed points by separatrices. Further separatrices connecting the saddle points at \(H > 0\) and \(H < 0\) are located on the plane \(K = 0\). These separatrices divide the phase space into four different regions. Within each region, trajectories emanate from one of the repulsive past finite time singularities and approach one of the attractive future finite time singularities. Hence, the universe is finite both in the past and the future. If the universe was initially contracting, one finds a bounce, after which it continues to expand.

\item
We then come to the case \(-\frac{1}{12} < \epsilon < 0\), which is shown in figure~\ref{fig:phase23m1_24} for the representative value \(\epsilon = -\frac{1}{24}\). Note that the location and properties of the projective fixed point~\eqref{eq:profix231} are unchanged, and will remain so for all values of \(\epsilon\). The saddle point~\eqref{eq:profix232} is now found near the bottom of the central vertical line \(K = 0\). The point~\eqref{eq:profix233} is now found to the left of the previously mentioned saddle point, and its stability is reversed combined to the previous case, i.e., it is now a repulsive past finite time singularity for \(H > 0\), and the opposite for \(H < 0\). In addition, we now see the appearance of further projective fixed points. One pair~\eqref{eq:profix234} is located on the central vertical line \(K = 0\), with the upper sign corresponding to a repeller for \(H > 0\), located near the top with \(L > 0\), while the lower sign represents an attractor near the bottom of the diagram at \(L < 0\). Finally, we find a pair~\eqref{eq:profix235} of saddle points, which are located at the left and at the right of the vertical line \(K = 0\).

The phase space is now divided by several separatrices connecting the various projective fixed points. From the discussion above that trajectories can emanate from four repellers in total, counting the projective fixed points on both hemispheres of the projected phase space, whose antipodes are therefore attractors. For the point~\eqref{eq:profix231}, we find that trajectories start from a past finite time singularity, and approach either one of the attractive, collapsing future finite time singularities given by the solution~\eqref{eq:profix233} or by the upper sign of the solution~\eqref{eq:profix234} with \(H < 0\), or the infinitely slowing expansion at the lower sign of the solution~\eqref{eq:profix234}, i.e., they approach any of the attractors except for the antipode of their point of origin. Trajectories starting from the past finite time singularity~\eqref{eq:profix233} can reach the attractors given by both signs of the solution~\eqref{eq:profix234}. Finally, trajectories emanating from the two repulsive projective fixed points~\eqref{eq:profix234} can reach any attractor except for their respective antipode. Several of these trajectories undergo a turnaround, passing from \(H > 0\) to \(H < 0\), while the opposite case of a bounce is not possible.

\item
A special case is given by \(\epsilon = -\frac{1}{12}\), which is shown in figure~\ref{fig:phase23m1_12}. Here we find that several projective fixed points coincide and their stability changes. This applies in particular to the points~\eqref{eq:profix232}, \eqref{eq:profix233} and the lower signs of the two solutions~\eqref{eq:profix234} and~\eqref{eq:profix235}, which are now all located at \(K = L + 3H = 0\) on the lower half \(L < 0\) of the central vertical line \(K = 0\) in the diagram. This common point now becomes a non-hyperbolic projective fixed point. There are only two repellers left, and trajectories emanating from these repellers as well as from the non-hyperbolic fixed points can reach either of the attractors (except for the antipode of the originating repeller). Trajectories which do not involve the non-hyperbolic fixed point come from a past finite time singularity and approach a future finite time singularity, so that the universe is finite, while trajectories involving the non-hyperbolic fixed point can also be infinite in either the past or the future, depending on the direction of their approach.

\item
We finally consider the case \(\epsilon < -\frac{1}{12}\), which we displayed using the choice \(\epsilon = -1\) in figure~\ref{fig:phase23m1}. Here we find that all projective fixed points are again distinct, and the stability of some of them has changed compared to the case \(-\frac{1}{12} < \epsilon < 0\). The point~\eqref{eq:profix233} has now become an attractor for \(H > 0\), and can be found in the lower right quarter of the diagram. The point given by the lower sign of the solution~\eqref{eq:profix234} has now become a repeller, and can be found as the lowest marked point in the upper half \(L > 0\) of the central vertical line \(K = 0\). Hence, we find again four repellers in the full projected phase space, whose antipodes are attractors. For the trajectories we find that now the trajectories emanating from each repeller can reach all attractors, except for the antipodes of their respective points of origin. This allows for trajectories which have all possible combination of finite and infinite past and future evolutions.
\end{enumerate}

\section{Conclusion}\label{sec:conclusion}
We studied the vacuum cosmological dynamics of the most general class of newer general relativity theories with a spatially flat Friedmann-Lemaître-Robertson-Walker metric and a homogeneous and isotropic symmetric teleparallel connection. For a particular subclass of theories, we find that the dynamical system becomes degenerate, and we derived the resulting constraints. Further, we studied the structure of the phase space, fixed points and effective dark energy. One of our most interesting findings is the fact that generically the vacuum cosmological dynamical system of newer general relativity assumes the homogeneous form \(\dot{\vek{z}} = \vek{f}(\vek{z}, \vek{z})\), where \(\vek{f}\) is a vector of constant symmetric bilinear forms, which is determined by the parameters of the theory. This insight allowed us to decompose the dynamics of the system into radial and angular parts, from which we obtained the asymptotic behavior and attractor solutions. In particular, we found that fixed points come in pairs with opposite stability properties depending on the sign of the Hubble parameter \(H\), and that there always exists a non-hyperbolic saddle point with \(H = 0\), which acts both as a sink and a source of trajectories.

We have then turned our attention to two particular classes of theories, which are distinguished by having the same post-Newtonian limit as general relativity and the absence of ghosts around a flat Minkowski. For the theories of type 1, we have found the complete set of vacuum cosmological solutions, and shown that none of them exhibits any accelerated expansion. We therefore conclude that these theories do not constitute a viable cosmological model, unless they are supplemented with other forms of dark energy to explain the observed late-time acceleration.

A richer and more interesting structure is found for theories of type 2. In this case we have seen that the vacuum cosmological dynamics crucially depend on the value of their single parameter \(\epsilon\). For \(\epsilon > 0\), we find that the effective dark energy barotropic index satisfies \(w_{\lambda} \leq -1\), and thus shows phantom behavior. For these models, there is no turnaround possible, and any expanding solution will inevitably lead to a future finite time singularity (big rip). Initially, these solutions originate from a complementary past finite time singularity for a contracting universe, which then undergoes a big bounce. In contrast, for \(\epsilon < 0\), we find an effective dark energy barotropic index \(w_{\lambda} \geq -1\), and thus non-phantom dark energy. As a consequence, we see that all solutions describing an expanding universe either lead to an eternal expansion, whose Hubble parameter asymptotically approaches \(0\), or undergo a turnaround to become contracting, and then lead to a future finite time singularity (big crunch).

It should be once again noted that in this work we have considered only vacuum solutions. While this is sufficient in order to determine attractor solutions for an expanding universe, in which the matter density can be neglected and the dynamics is dominated by the symmetric teleparallel geometry, which takes the role of dark energy in these theories as the driving force of an accelerating expansion, a full and comprehensive analysis of the cosmological history must also take matter into account. Due to the complexity and vast number of cosmological solutions already in the vacuum case, we defer this full analysis to future work.

Another future line of research is to study cosmological perturbations around the solutions we have found, using the methods outlined in~\cite{Heisenberg:2023wgk,Heisenberg:2023tho}. This is in particular relevant since cosmological perturbations in symmetric teleparallel gravity have been found to be generically pathological, leading to a strong coupling problem or ghost instabilities~\cite{BeltranJimenez:2019tme,BeltranJimenez:2021auj,Gomes:2023tur}. A possible solution to these issues could be found by considering cosmological models of lower symmetry, in which the universe is assumed to be only homogeneous, but not isotropic, which would constitute a class of symmetric teleparallel Bianchi spacetime models.

\begin{acknowledgments}
This research was funded by the Science Committee of the Ministry of Science and Higher Education of the Republic of Kazakhstan (Grant No. AP26101851).
MH gratefully acknowledges the full financial support by the Estonian Research Council through the Personal Research Funding project PRG2608 and the Center of Excellence TK202 ``Fundamental Universe''.
\end{acknowledgments}

\bibliography{nmdynsys}

\end{document}